\documentclass[twocolumn]{aastex631}
\usepackage{amsmath}
\usepackage[T1]{fontenc}
\usepackage{upgreek}

\newcommand{\PSUAA}{Department of Astronomy \& Astrophysics, 525 Davey Laboratory, The Pennsylvania State University, University Park, PA, 16802, USA}
\newcommand{\PSUCEHW}{Center for Exoplanets and Habitable Worlds, 525 Davey Laboratory, The Pennsylvania State University, University Park, PA, 16802, USA}

\newcommand{\comment}[1]{}

\newcommand{\rrev}{}

\newcommand{\rreva}{}

\submitjournal{AJ}

\shorttitle{EPRV Survey Simulations}
\shortauthors{Gupta \& Bedell}

\graphicspath{{./}{figures/}}

\begin{document}

\title{Fishing for Planets: A Comparative Analysis of EPRV Survey Performance in the Presence of Correlated Noise}

\correspondingauthor{Arvind F.\ Gupta}
\email{arvind.gupta@noirlab.edu}

\author[0000-0002-5463-9980]{Arvind F.\ Gupta}
\affil{\PSUAA}
\affil{\PSUCEHW}

\author[0000-0001-9907-7742]{Megan Bedell}
\affil{Center for Computational Astrophysics, Flatiron Institute, 162 5th Avenue, New York, NY 10010, USA}

\begin{abstract}

With dedicated exoplanet surveys underway for multiple extreme precision radial velocity (EPRV) instruments, the near-future prospects of RV exoplanet science are promising.
These surveys' generous time allocations are expected to facilitate the discovery of Earth analogs around bright, nearby Sun-like stars.
But survey success will depend critically on the choice of observing strategy, which will determine the survey's ability to mitigate known sources of noise and extract low-amplitude exoplanet signals.
Here, we present an analysis of the Fisher information content of simulated EPRV surveys, accounting for the most recent advances in our understanding of stellar variability on both short and long timescales (i.e., oscillations and granulation within individual nights, and activity-induced variations across multiple nights).
In this analysis, we capture the correlated nature of stellar variability by parameterizing these signals with Gaussian Process kernels.
We describe the underlying simulation framework as well as the physical interpretation of the Fisher information content, and we evaluate the efficacy of EPRV survey strategies that have been presented in the literature. We explore and compare strategies for scheduling observations over various timescales and we make recommendations to optimize survey performance for the detection of Earth-like exoplanets.

\end{abstract}

\section{Introduction}\label{sec:fisher_introduction}

For several decades, astronomers have been carrying out systematic radial velocity (RV) surveys to search for exoplanets around nearby stars. The first detection of an exoplanet orbiting a main sequence star by \citet{Mayor1995} was the result of one such search, and this discovery spurred decades of interest and accelerating progress in exoplanet science.
As instrumentation and analysis techniques have improved, as evidenced by the growing class of spectrographs capable measuring Doppler signals with sub-m s$^{-1}$ precision, so too have the prospects of exoplanet surveys.

Several of these extreme precision radial velocity (EPRV) spectrographs support exoplanet surveys with generous time allocations, totalling hundreds to thousands of hours each year for up to a decade. 
These include 
the EXPRES 100 Earths Survey \citep{Jurgenson2016,Brewer2020},
the NEID Earth Twin Survey \citep{Schwab2016,Gupta2021},
the upcoming Terra Hunting Experiment \citep{Hall2018} with HARPS-3 \citep{Thompson2016},
the HARPS-N Rocky Planet Search \citep{Cosentino2012,Motalebi2015},
and a blind radial velocity survey with ESPRESSO \citep{Pepe2021,Hojjatpanah2019}.
While the exact objectives of each survey vary, two common themes are the discovery and characterization of habitable-zone, Earth-mass exoplanets and the refinement of intrinsic exoplanet occurrence rate statistics, particularly for exoplanets in mass and period regimes that were inaccessible to previous instruments.

\rrev{
Even with substantial time allocations and precise instruments, detecting sub-m~s$^{-1}$ signals will be a challenge. In this regime, RV uncertainties are dominated by correlated noise from intrinsic stellar variability, which has been identified as the largest remaining hurdle for RV detection of Earth-like exoplanets \citep{Crass2021}. We expect our ability to mitigate these correlated noise contributions and extract exoplanet-induced signals will be greatly impacted by the enacted observing scheme \citep{Dumusque2011,Hall2018,Chaplin2019,Luhn2023}. Indeed, recent studies on the scheduling of RV observations to confirm and characterize transiting systems \citep{Burt2018,Cloutier2018,Cabona2021} as well as for large-scale planet searches \citep{Luhn2023} have shown that the choice of observing strategy will impact survey efficacy in the presence of correlated noise. To identify an optimal strategy for Earth analog searches, a sizeable parameter space of survey strategies may need to be explored.
}

\rreva{Various methods have been used to assess the efficiency of exoplanet detection with RV searches, including Bayesian inference for adaptive schedules \citep{Ford2008,Loredo2011} and injection and recovery of planetary and stellar signals \citep[e.g.,][]{Hall2018}. Here, } we \rrev{follow \citet{Baluev2008} and \citet{Baluev2009} and we use Fisher information analysis to assess and compare RV exoplanet search strategies.}

In this work, we present a framework in which Fisher information \rrev{is used} to quantify the intrinsic information content of simulated RV observations and to predict the detection limits achieved by ongoing and future EPRV exoplanet surveys in a computationally efficient manner.
We show that such a tool can be used to identify survey strategies that will optimize sensitivity to specific populations of exoplanets, such as Earth analogs.
We describe the Fisher information calculation in Section \ref{sec:fisher} and the input Gaussian process (GP) kernels with which we parameterize contributions from stellar variability in Section \ref{sec:fisher_covariance}.
In Section \ref{sec:unit_tests}, we compare the detection limits achieved by sets of simulated observations across various timescales to formulate an idealized survey design. 
We apply these findings to full survey simulations in Section \ref{sec:fisher_simulations}, in which we also describe our observation simulation tool that accounts for practical observing constraints to ensure realistic survey realizations.
And we discuss our results and their implications in Section \ref{sec:fisher_results} and caveats and prospects for future work in Section \ref{sec:fisher_discussion}.

\section{Fisher Information}\label{sec:fisher}

The Fisher information content of a  data set can be used to calculate the expected uncertainty on a set of parameters for a representative model \citep{Fisher1922}. The Fisher information \rrev{matrix}, $B$, is given as the expectation value  of the Hessian of the negative log likelihood function \rreva{taken over data $x$}
\begin{equation}
    B_{i,j} = -E_x\left[\frac{\partial^2\ln \mathcal{L}(x;\theta)}{\partial \theta_i \partial \theta_j}\right]
\end{equation}
where $\theta$ is the parameter vector and the log likelihood is
\begin{equation}
\begin{split}
    {\rm ln}\ \mathcal{L} = -\frac{1}{2}&\left[(x-\mu)^T C^{-1} (x-\mu)\right.\vphantom{.} \\ &+ \left.\vphantom{.} N\ {\rm ln}\ 2\pi + {\rm ln\ det}\ C\right]
\end{split}
\end{equation}
for a model $\mu$ and time series $x$ of length $N$, and a $N\times N$ covariance matrix, $C$, that describes the expected noise properties of the measurements. 
Both $\mu$ and $C$ may depend on $\theta$ in the general case.

\rrev{Assuming that the data are drawn from a multivariate Gaussian distribution with mean $\mu$ and covariance $C$}, the Fisher information can be written as
\begin{equation}
\begin{split}
    B_{i,j} = & \left(\frac{\partial \mu}{\partial \theta_i}\right)^TC^{-1}\left(\frac{\partial \mu}{\partial \theta_j}\right)\\ & + \frac{1}{2}{\rm tr}\left(C^{-1}\frac{\partial C}{\partial \theta_i}C^{-1}\frac{\partial C}{\partial \theta_i}\right).
\end{split}
\end{equation}
\rrev{For a derivation, see \citet{Kay1993} or \citet{Malago2015}.} If the covariance is independent of the model parameters, \rrev{this reduces to}
\begin{equation}\label{eq:fisher}
    B_{i,j} = \left(\frac{\partial \mu}{\partial \theta_i}\right)^TC^{-1}\left(\frac{\partial \mu}{\partial \theta_j}\right).
\end{equation}
The diagonal elements of the inverse of the Fisher information matrix represent the parameter uncertainties:
\begin{equation}\label{eq:fisher_sig}
    \sigma_{\theta_i}^2 = B_{i,i}^{-1}.
\end{equation}
It is important to note here that the time series $x$ has dropped out of the equation \rrev{because of the average taken over this value}, and the Fisher information depends only on the model $\mu$, the covariance matrix $C$, and the times $t$ at which measurements are taken. That is, we do not need to know the measured values of $x$ to determine the expected parameter uncertainties.

The use of Fisher information analyses for astronomical data sets was first studied by \citet{Tegmark1997}, who explored the utility of this metric in the context of cosmology\rrev{, and later works applied Fisher information to the optimization of observing strategies for RV exoplanet science \citep[e.g.,][]{Baluev2008,Baluev2009,Baluev2013}.}
More recent applications include the work of \citet{Gomes2022}, who assess sensitivity to gravitational effects of an as-yet undetected outer Solar System planet, and that of \citet{Cloutier2018}, who \rrev{used Fisher information to explore the impact of correlated noise signals on follow-up observations of transiting exoplanets.}  Here, we adopt the notation used in \citet{Cloutier2018} and present a framework for using Fisher information to calculate detection sensitivity limits for blind RV exoplanet surveys. The RV semi-amplitude induced by a single exoplanet with mass $M_p$ and period $P$ orbiting a star with mass $M_\star$  is
\begin{equation}\label{eq:semiamp}
    K =  \frac{M_p \sin i}{\sqrt{1-e^2}}  \left(\frac{2\pi G}{(M_\star + M_p)^2P}\right)^{1/3}.
\end{equation}
For our model in this work, we assume circular, zero-eccentricity orbits and single-planet systems such that the exoplanet-induced RV signal can be represented with a simplified Keplerian
\begin{equation}\label{eq:rvmodel}
    \mu(t) = K \sin \left(\frac{2\pi}{P} t -\phi_0\right),
\end{equation}
where $\phi_0$ is a phase offset and $t$ is the set of observation times. The corresponding parameter vector is then
\begin{equation}
    \theta = \{K, P, \phi_0\},
\end{equation}
and the derivative of the model with respect to this vector is
\begin{equation}
    \frac{\partial\mu}{\partial\theta} =
        \begin{bmatrix}
            \frac{\partial\mu}{\partial K} \\ \frac{\partial\mu}{\partial P} \\ \frac{\partial\mu}{\partial\phi_0}
        \end{bmatrix} =
        \begin{bmatrix}
            \sin(\frac{2\pi}{P}t - \phi_0) \\ -\frac{2\pi t}{P^2} K\cos(\frac{2\pi}{P}t - \phi_0) \\ -K\cos(\frac{2\pi}{P}t - \phi_0)
        \end{bmatrix}.
\end{equation}
The inclusion of a covariance matrix in \autoref{eq:fisher} allows us to capture the effect of correlated noise on the achieved RV sensitivity.
In a pure white noise scenario, $C$ will be a matrix with elements $C_{n,m} = \sigma_n^2\delta_{nm}$, where $\sigma_n$ is the total measurement uncertainty for observation $1\leq n \leq N$ and $\delta$ is the delta function
\begin{equation}
\delta_{nm} =
\begin{cases} 
      1 & m=n \\
      0 & m\neq n 
   \end{cases}.
\end{equation}
That is, all the off-diagonal elements will be $0$; independent measurements will have independent noise properties.
But if correlated noise sources are present, some off-diagonal elements will be non-zero as well, representing covariances between the noise properties of pairs of observations.
We discuss anticipated sources of noise and the construction of a covariance matrix for EPRV data sets in the following section.


\section{Constructing a Covariance Matrix for Radial Velocity Observations}\label{sec:fisher_covariance}

The full covariance matrix to be used in \autoref{eq:fisher} will be made up of a sum of white noise and correlated noise components
\begin{equation}\label{eq:covar}
    C = \sigma_{\rm photon}^2I +k_{\rm osc}+k_{\rm gran}+k_{\rm activity},
\end{equation}
where $I$ is the $N\times N$ identity matrix and the remaining terms are described below.

The primary source of white noise is photon noise, or the limit on the precision of an individual RV measurement as determined by the Doppler information content of the observed stellar spectrum. The photon noise limit, $\sigma_{\rm photon}$, depends on the wavelength range and spectral features from which the RV measurement is derived and on the observed signal-to-noise ratio (S/N) across this spectral range, and it can be calculated analytically as shown by \citet{Bouchy2001}.

Instrument systematics and RV variations due to intrinsic stellar variability constitute two important sources of correlated noise that will impact RV exoplanet discovery efforts \citep{Crass2021}. 
In this work, we will focus primarily on the correlated nature of stellar variability.
Well-characterized instrument systematics are typically modeled and calibrated out during the data reduction process for EPRV spectrographs \citep[e.g.,][]{Halverson2016,Petersburg2020}. Though systematics persist at a level that obstructs the detection of 10 cm~s$^{-1}$ level signals, progress on calibration and reduction techniques \rreva{\citep[e.g.,][]{Zhao2021,Cretignier2023}} may make them manageable in the near future. Here, we consider the post-calibration RV time series as our starting point, such that any systematics that can be removed have been removed, and we will reserve a more detailed discussion of instrument-related correlated noise for Section \ref{sec:fisher_instruments}.

Intrinsic stellar RV variations are present in the final RV time series at the $\sim1$ m~s$^{-1}$ level for typical observations of quiet exoplanet search targets.
To account for these stellar signals, it is common practice to model them using Gaussian process (GP) regression while simultaneously fitting for an exoplanet-induced signal \citep[see review by][]{Aigrain2022}. This approach provides for straightforward integration with the covariance matrix, as one can build individual sources of variability into this matrix with appropriate GP kernel functions. In the remainder of this section, we describe the dominant sources of stellar variability for typical Sun-like G- and K-dwarfs and we present the form of their respective GP kernels.

On short timescales, i.e., less than a single night, the most important sources of stellar RV variations are p-mode oscillations, which manifest as m~s$^{-1}$-level variations on 5-10 minute timescales, and surface granulation, which has a similar amplitude on slightly longer timescales.
For oscillations, we adopt the GP kernel described by \citet{Luhn2023} and based on work by \citet{Pereira2019} and \citet{Guo2022}, which takes the form
\begin{equation}\label{eq:kosc}
\begin{split}
    k_{\rm osc}(\Delta) & = S_{\rm osc} \omega_{\rm osc} Q e^{\frac{-\omega_{\rm osc}\Delta}{2Q}}\\
    & \times \left(\cos (\eta \omega_{\rm osc}\Delta) + \frac{1}{2\eta Q}\sin (\eta \omega_{\rm osc} \Delta)\right),
\end{split}
\end{equation}
where $\Delta$ is the $N\times N$ matrix that represents the absolute value of the time delay between pairs of observations
\begin{equation}
    \Delta_{n,m} = |t_n - t_m|.
\end{equation}
$S_{\rm osc}$ is the power at the peak of the oscillation excess, $\omega_{\rm osc}$ is the characteristic oscillation frequency, $Q$ is the quality factor, and $\eta= |1-(4Q^2)^{-1}|^{1/2}$.

\begin{figure*} 
    \centering
    \includegraphics[width=1.0\linewidth]{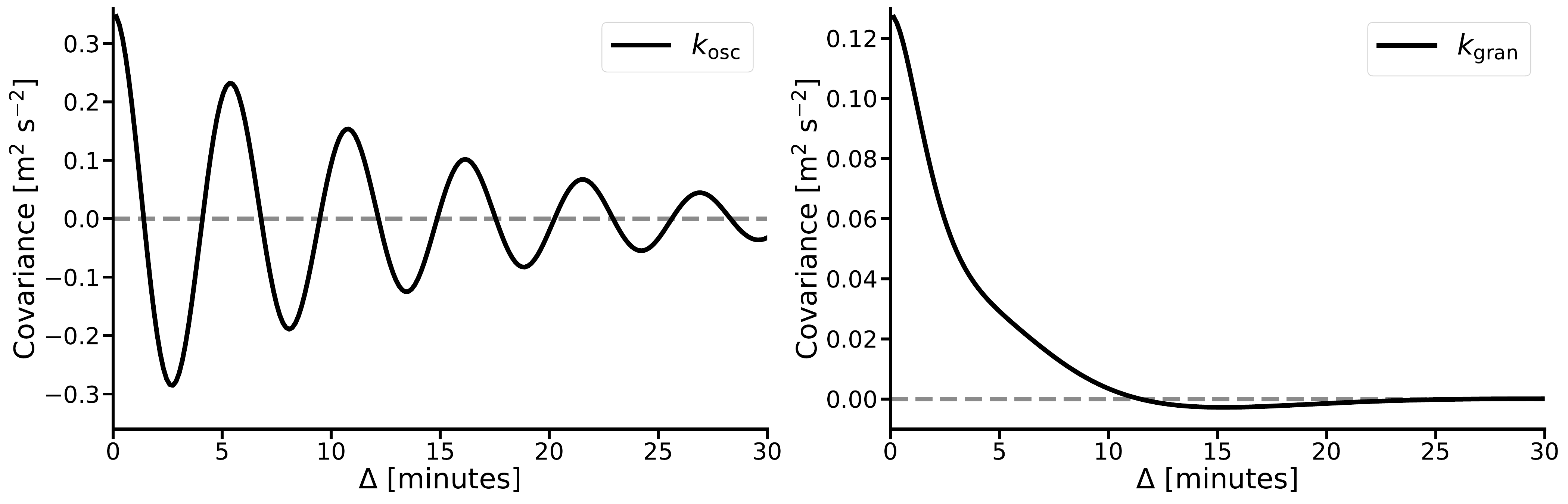}
    \caption{Covariance between pairs of observations as a function of separation in time for the GP kernels for stellar oscillations (\autoref{eq:kosc}; left) and for stellar surface granulation (\autoref{eq:kgran}; right). We assume instantaneous exposures for both kernels. We show that oscillations and granulation only introduce significant RV noise on short timescales of less than a single night.}
    \label{fig:fisher_covar_short}
\end{figure*}

We also adopt the following two-component granulation kernel from \citet{Luhn2023} and \citet{Guo2022}:
\begin{equation}\label{eq:kgran}
\begin{split}
    k_{\rm gran}(\Delta) & = S_1\omega_1e^{\frac{-\omega_1 \Delta}{\sqrt{2}}}\cos\left(\frac{\omega_1\Delta}{\sqrt{2}}-\frac{\pi}{4}\right)\\
    & + S_2\omega_2e^{\frac{-\omega_2 \Delta}{\sqrt{2}}}\cos\left(\frac{\omega_2\Delta}{\sqrt{2}}-\frac{\pi}{4}\right),
\end{split}
\end{equation}
where $S_1$ and $S_2$ describe the power of each granulation component and $\omega_1$ and $\omega_2$ are the respective characteristic frequencies.
\rrev{
We note that our granulation model does not account for supergranulation, which typically occurs on timescales of longer than a day. Supergranulation is not a negligible source of stellar RV variability \citep[e.g.,][]{Meunier2015,AlMoulla2023}, but we do not include it in this work because we do not have a suitable model.
}
We show the form of the oscillation and granulation kernels in \autoref{fig:fisher_covar_short}, assuming perfectly Solar hyperparameters as given in \autoref{tab:fisher_gp_hyper}.

\begin{deluxetable}{llrl}
\tablecaption{Solar Hyperparameter Values for Stellar Variability Gaussian Process Kernels \label{tab:fisher_gp_hyper}}
\tablehead{\colhead{Kernel}&  \colhead{Parameter}&  \colhead{Value}&  \colhead{Units}}
\startdata
$^{\rm a}k_{\rm osc}$ & $S_{\rm osc}$  & $2.36$& m$^2$ rad$^{-1}$ s$^{-1}$\\
&$\omega_{\rm osc}$  & $0.0195$& rad s$^{-1}$\\
&$Q$ & $7.63$ &\\
$^{\rm a}k_{\rm gran}$  & $S_1$ & $18.64$ & m$^2$ rad$^{-1}$ s$^{-1}$\\
&$S_2$ & $5.22$ & m$^2$ rad $^{-1}$ s$^{-1}$ \\
&$\omega_1$  & $0.00485$ & rad s$^{-1}$\\
&$\omega_2$ & $0.0173$ & rad s$^{-1}$\\
$^{\rm b}k_{{\rm activity}, QP}$ &$\alpha$  & $1.44$ & m s$^{-1}$\\
&$\Gamma$ &$1.486$ & \\
&$\lambda_1$  & $23.6$ & days\\
&$\lambda_2$  & $28.1$ & days\\
$^{\rm c}k_{{\rm activity}, M\,52}$ & $a_0$  &  $-0.106884$ & m s$^{-1}$\\
& $a_1$  & $-1.154$ & m s$^{-1}$\\
& $\lambda$  & $2.52$ & days\\
\enddata
\tablenotetext{{\rm a}}{Parameters calculated following the precsription of \citet{Luhn2023} for $T_{\rm eff} = 5777$ K, $\log g$ = 4.43, $\nu_{\rm max} = 3100$ $\upmu$Hz}
\tablenotetext{{\rm b}}{Parameters derived by \cite{Langellier2021} via fit to HARPS-N solar data}
\tablenotetext{{\rm c}}{Parameters given by \citet{Luhn2023}, derived from \citet{Gilbertson2020b} spot-induced solar RV variability simulations}
\end{deluxetable}

\begin{figure*} 
    \centering
    \includegraphics[width=1.0\linewidth]{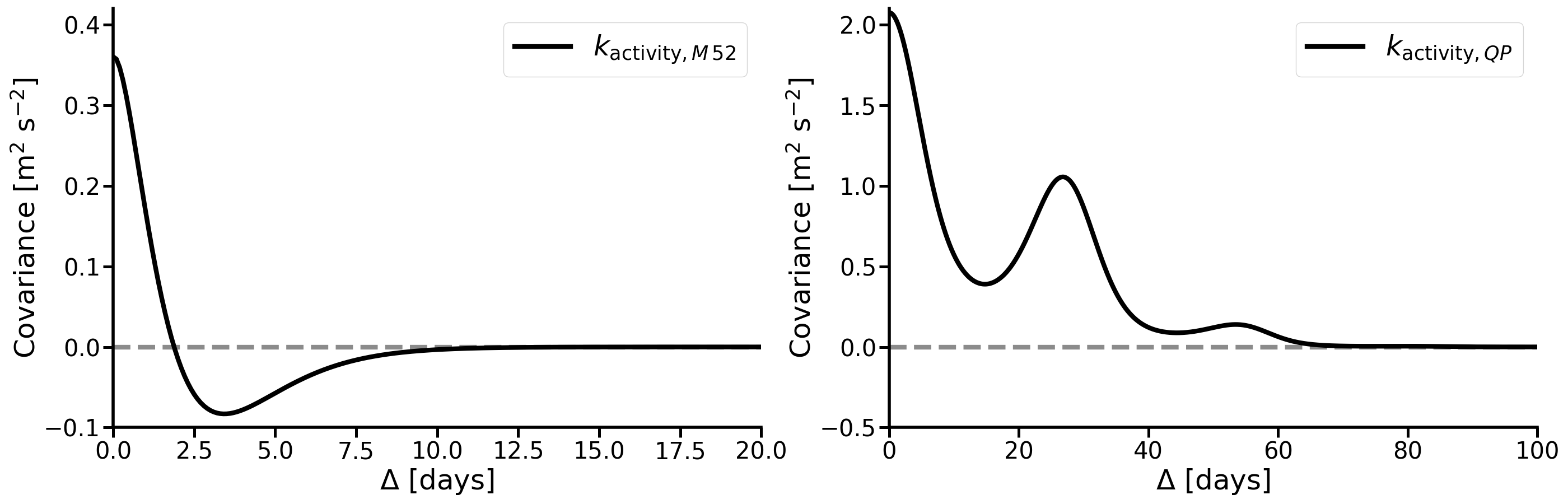}
    \caption{Covariance between pairs of observations as a function of separation in time for the quasiperiodic kernel for rotationally-modulated stellar activity (\autoref{eq:qp_act}; left)  and for the Mat\'ern-5/2 kernel for rotationally-modulated stellar activity (\autoref{eq:mat_act}; right). While the covariance for both kernels persists across many nights, we note that the curves differ significantly in amplitude and shape.}
    \label{fig:fisher_covar_long}
\end{figure*}

Stellar RV variability resulting from rotational modulation of active regions and starspots is our primary source of correlated noise on longer timescales. Activity-induced variability is commonly modeled using a quasi-periodic GP kernel of the form
\begin{equation}\label{eq:qp_act}
    k_{{\rm activity}, QP}(\Delta) = \alpha^2 \exp\left(-\frac{\Delta^2}{2\lambda_1^2}-\Gamma \sin^2\left(\frac{\pi\Delta}{\lambda_2}\right)\right)
\end{equation}
where $\alpha$ is the RV amplitude, $\lambda_1$ is the coherence timescale, $\lambda_2$ represents the stellar rotation period, and $\Gamma$ is the periodic complexity factor, or in this case, the complexity of the spot and active region coverage across the stellar surface \rrev{\citep{Aigrain2012,Haywood2014}}.
This kernel has been shown to be effective for disentangling exoplanet- and activity-induced RV signals, and each hyperparameter has a clear physical interpretation, making it a natural choice for modeling RV time series data in the presence of activity. However, recent work by \citet{Gilbertson2020b} based on simulated solar spectra \rrev{with short-lived spots} \citep{Gilbertson2020a} has shown other kernels outperform the quasi-periodic kernel, particularly for detecting exoplanets with $K\leq30$ cm~s$^{-1}$. We therefore also consider the following latent GP kernel
\begin{equation}\label{eq:mat_act}
    k_{{\rm activity}, M\,52}(\Delta) = a_0^2 k_{M\,52}(\Delta) -a_1^2\frac{d^2}{dt^2} k_{M\,52}(\Delta)
\end{equation}
which is a linear combination of the Mat\'ern-5/2 kernel
\begin{equation}
    k_{M\,52}(\Delta) = e^{-\frac{\sqrt{5}\Delta}{\lambda}}\left(1+\frac{\sqrt{5}\Delta}{\lambda}+\frac{5\Delta^2}{3\lambda^2}\right)
\end{equation}
and its second derivative, with a single timescale hyperparameter, $\lambda$, and amplitude coefficients $a_0$ and $a_1$. We refer the reader to \citet{Gilbertson2020b} and \citet{Luhn2023} for the calculation of the second derivative of the Mat\'ern-5/2 kernel and for a full derivation of \autoref{eq:mat_act}. We show both activity kernels in \autoref{fig:fisher_covar_long}. For the Mat\'ern-5/2 kernel, we assume hyperparameters as derived by \citet{Gilbertson2020b} and for the quasiperiodic kernel, we assume hyperparameters from a fit to HARPS-N solar data by \citep{Langellier2021}. Both sets of hyperparameters are given in \autoref{tab:fisher_gp_hyper}. In subsequent sections, we test survey performance using each of the two activity kernels and comment on how our conclusions change based on the assumed kernel form.

\section{Formulating an Idealized Survey Strategy}\label{sec:unit_tests}

We begin our investigation of survey performance with a set of experiments intended to isolate each source of correlated noise and identify observing strategies that will best mitigate their respective impacts on exoplanet detection sensitivity.
Separate sets of simulated observing schedules are generated to highlight the effects of stellar RV signals that vary on timescales of less than a single night for Sun-like stars, granulation and p-mode oscillations, and those that span multiple nights, namely active regions and spots.
For each set of observations, we calculate the resulting Fisher information and assess the achieved sensitivity limits. In all cases, we adopt perfectly Solar hyperparameters for the GP kernels with which we build each covariance matrix. \rrev{We assume that only a single planet is present in the system and we take the stellar noise model to be fixed.}

\subsection{Intra-night Observing Schedule Simulations}\label{sec:intranight_sim}

As we show in \autoref{fig:fisher_covar_short}, correlations in oscillation and granulation RV signals are not expected to remain coherent for longer than an hour for Sun-like stars, so observations collected on separate nights are effectively independent. 
To assess the impact of oscillations and granulation on detection sensitivity limits, we therefore explore different strategies for distributing observations within a single night. We consider the following intra-night cadences: 
\begin{itemize}
    \item six visits with one exposure per visit,
    \item three visits with two exposures per visit,
    \item two visits with three exposures per visit,
    \item one visit with six exposures,
    \item and one visit with a single continuous exposure.
\end{itemize}
For each of these cadences, we generate 100 nights of observation times across a 500-night baseline (average inter-night separation of 5 nights), and we repeat this for exposure times $15$ s $\leq t_{\rm exp}\leq1200$ s. For the three strategies with multiple visits, the start times of each visit are evenly distributed across a six hour window. For the three strategies with multiple exposures in a single visit, the exposures are separated by a readout time, $t_r$. We then calculate the Fisher information of the simulated observations for covariance matrices consisting of correlated noise from (i) p-mode oscillations, (ii) granulation, (iii) p-mode oscillations and granulation, and (iv) p-mode oscillations, granulation, and photon noise. We do not include an instrumental jitter term for these tests.

For oscillations and granulation, we start with the GP kernels in Equations \ref{eq:kosc} and \ref{eq:kgran}. However, as \citet{Luhn2023} note, exposures are not instantaneous, and the exposure times we consider here are comparable to the  variability timescales for oscillations and granulation in the case of Sun-like stars. We therefore compute the double integral of these GP kernels over the exposure times to calculate the true covariance between pairs of observations. A detailed discussion of the oscillation and granulation kernel integration is given in the Appendix of \citet{Luhn2023}. The assumed hyperparameters are given in \autoref{tab:fisher_gp_hyper}. We determine the photon noise using the NEID exposure time calculator, described in \citet{Gupta2021}, for Solar analogs with $V = 4$ and 8 mag.

\begin{figure*} 
    \centering
    \includegraphics[width=1.0\linewidth]{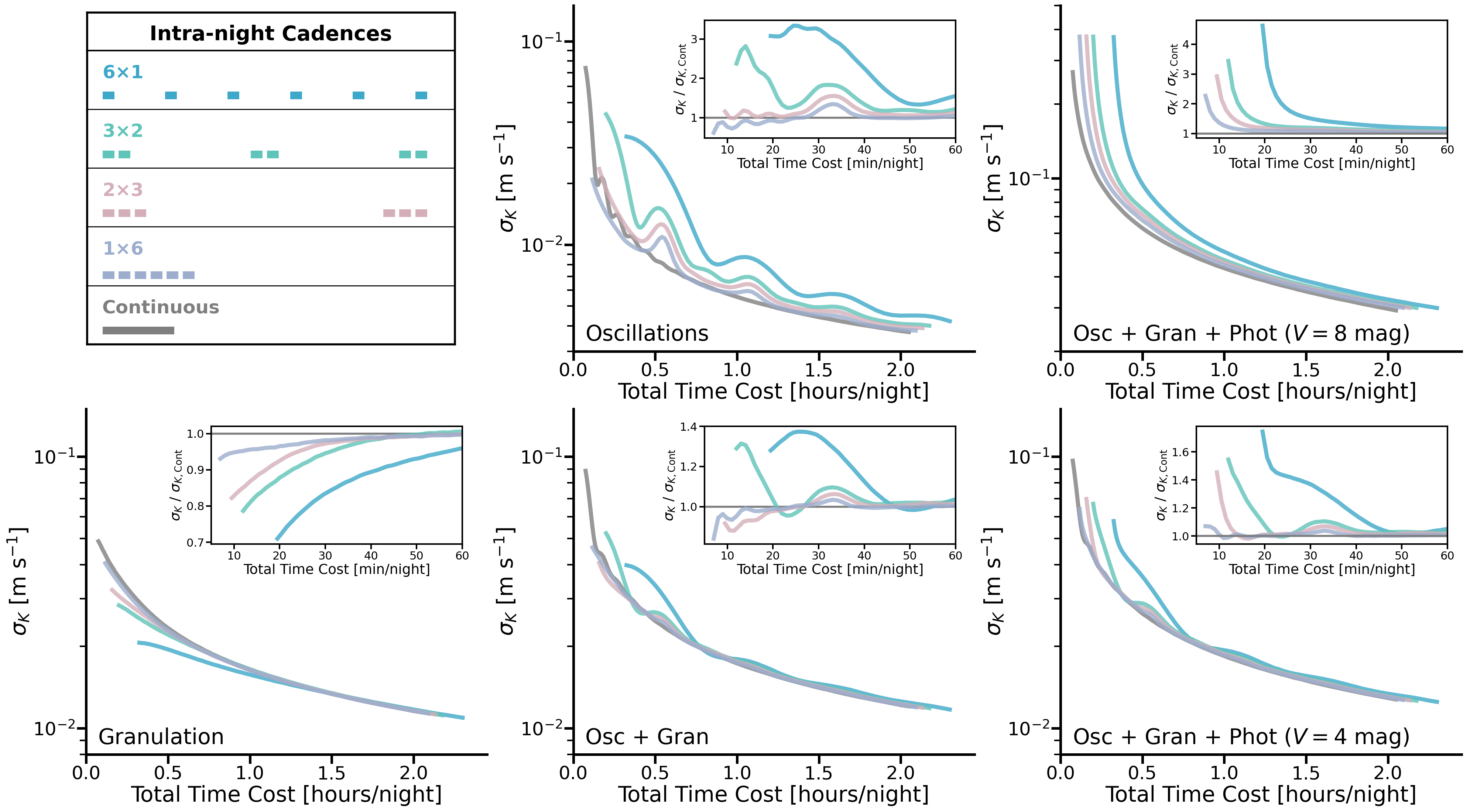}
    \caption{Expected semi-amplitude uncertainty, $\sigma_K$, for a $K=10$ cm~s$^{-1}$ signal and an orbital period of $P=100$ days as a function of total nightly time cost (as calculated via \autoref{eq:tnight}). We calculate $\sigma_K$ for simulated observing schedules with 100 nights of observations distributed across 500 total nights, wherein the observations on each night are distributed according to each of the five strategies depicted in the upper left panel. These strategies are described in Section \ref{sec:intranight_sim}. In the remaining panels, we show the impact of contributions from various combinations of photon noise and correlated noise from oscillations and granulation. We also include inset plots to highlight the performance of each multi-exposure strategy relative to the single continuous exposure strategy for nightly time costs of less than an hour.}
    \label{fig:fisher_oscgran}
\end{figure*}

Finally, we use the covariance matrices and simulated observation times to evaluate the performance of each strategy, where the performance is defined as the expected uncertainty on the semi-amplitude, $\sigma_K$, using Equations \ref{eq:fisher} and \ref{eq:fisher_sig}. We calculate this value for an exoplanet with orbital parameters $K=10$ cm~s$^{-1}$ and $P=100$ d. \rrev{To marginalize over orbital phase, we perform the calculation for 10 values of $\phi_0$} uniformly distributed on the interval $0\leq\phi_0<2\pi$ and we use the mean value of the \rreva{resulting Fisher information uncertainty estimates}. The dependence of $\sigma_K$ on the total nightly time cost of each strategy is shown in Figure \ref{fig:fisher_oscgran}, where the time cost is calculated as the sum of the exposure times and associated overheads, i.e., readout and target acquisition time ($t_{\rm acq}$):
\begin{equation}\label{eq:tnight}
    t_{\rm night} = N_{\rm visits}(t_{\rm acq} + N_{\rm exp}t_{\rm exp}+(N_{\exp}-1)t_r).
\end{equation}
We assume here that the target need only be acquired once per visit and that the readout cost of the final exposure in a sequence can \rrev{be} folded into the cost of acquiring the subsequent target. We use overhead costs of 30 seconds and 180 seconds for readout and acquisition, respectively, to match typical overheads for observations of bright stars with the NEID spectrograph.

\subsection{Intra-night Observing Strategy Results}\label{sec:intranight_res}

In \autoref{fig:fisher_oscgran}, we show that in the presence of correlated noise from oscillations alone, the expected semi-amplitude uncertainty, $\sigma_K$, falls off much more rapidly for the single-visit strategies than for the multi-visit strategies across all nightly time allocations. These strategies take advantage of the significant negative covariance seen on timescales comparable to $1/\nu_{\rm max}$, rapidly averaging out the p-mode signal rather than allowing it to bin down like white noise when the covariance approaches zero (i.e., when exposures are widely separated relative to the oscillation periods for Sun-like stars). But it is interesting that neither $\sigma_K$ nor the relative performance of the different intra-night strategies changes monotonically with nightly time cost. Instead, each intra-night strategy produces several local minima and maxima in $\sigma_K$, and these features emerge at different values of $t_{\rm night}$. For granulation, on the other hand, while $\sigma_K$ does decrease monotonically, the relative performance of each observing strategy is inverted. This can be attributed to the positive covariance of the granulation signal, which is not conveniently averaged out on short timescales. In contrast to oscillations, the correlated noise contribution from granulation is optimally minimized by taking exposures separated by $\Delta\gtrsim15$ minutes, where the covariance asymptotically approaches zero.
However, our results for the combined case with both oscillations and granulation do not consistently prefer the six-visit strategy. Instead, every strategy yields the lowest value of $\sigma_K$ for at least one value of $t_{\rm night}$ when we include both covariance kernels.
That is, the relative advantages of a single-visit strategy, which is optimal for averaging out the oscillation signal, and a multi-visit strategy, which is optimal for binning down the granulation signal, will win out on different time scales. It is also important to note that the optimal strategy for any given nightly time cost will depend on the oscillation and granulation kernel hyperparameter of the specific star being observed. Here, we have assumed \rrev{the hyperparameters are fixed at the Solar values}, but these will vary from star to star.

These results are largely consistent with those of previous studies that explore the RV noise contributions from oscillations and granulation. \citet{Chaplin2019} recently presented a model for predicting residual p-mode oscillation amplitudes as a function of exposure time and stellar parameters, showing that local minima in the residual amplitude curve are present for exposure times close to integer multiples of $\nu_{\rm max}$, the central frequency of the oscillation power excess.
\rrev{Their predictive }model has since become widely used in the design of EPRV observations, both for exoplanet surveys \citep[e.g.,][]{Brewer2020,Gupta2021} as well as for studies of stellar signals \citep[e.g.,][]{Sulis2023,Gupta2022}. The covariance behavior of the oscillation kernel we use here (\autoref{eq:kosc}) produces these same local minima, albeit on different timescales due to our inclusion of overheads costs and noncontiguous observations. But while the \citet{Chaplin2019} model \rreva{highlights the merits of a mitigation strategy} that averages over the oscillations, our results demonstrate the relative advantage of sampling the oscillation signal instead. In the Fisher information framework, the strategy with six consecutive exposures outperforms the continuous exposure strategy on short timescales, because this sampling allows us to use the measured RVs and known GP kernel form to model out the oscillation signal.

We also compare our findings to those of \citet{Dumusque2011}, who simulate measurements of stellar RV signals from fitted asteroseismic power spectra, including both oscillations and granulation, and calculate the predicted RV rms for several years of simulated data with different intra-night exposure times and cadences. For the Sun-like, G2V star $\alpha$ Cen A, \citet{Dumusque2011} find that they can achieve a lower RV rms by distributing observations across many hours instead of concentrating them in a single visit with the same total on-sky exposure time. We see the same trend for the granulation-only case in \autoref{fig:fisher_oscgran}, in which the six-visit strategy outperforms the other strategies by a significant margin for nightly time costs up to one hour. \rrev{However, our results differ from theirs for the case with both oscillations and granulation. We attribute this to two factors.} First, \citet{Dumusque2011} do not account for overheads when comparing the performances of different observing cadences. Cadences with more visits are not penalized for the additional costs that will be incurred, so their performance relative to cadences with fewer visits \rrev{appears enhanced}. On the other hand, their study includes the effects of supergranulation and a constant instrument noise floor, both of which introduce correlated noise on timescales longer than oscillations and granulation alone. Because we neglect these sources of correlated noise, we diminish the relative benefits of increasing the spacing between observations. We comment on correlated noise from instrument systematics in more detail in Section \ref{sec:fisher_instruments}. \rrev{We also note that other differences between the \citet{Dumusque2011} methodologies and our own, such as the precise oscillation and granulation models used, may have an effect on the differences between the results.}

The final noise source we consider here is photon noise. In \autoref{fig:fisher_oscgran}, we see that the multi-visit strategies are penalized more heavily due to the additional overhead costs. That is, for a given time allocation, increasing the number of visits will decrease the total exposure time and thus degrade the RV precision and achieved uncertainty on $K$. The single-visit strategies are again preferred across most nightly time costs. This preference is stronger for fainter stars, for which photons noise contributes a larger share of the total uncertainty.

\subsection{Inter-night Observing Schedule Simulations}\label{sec:internight_sim}

\begin{figure} 
    \centering
    \includegraphics[width=1.0\linewidth]{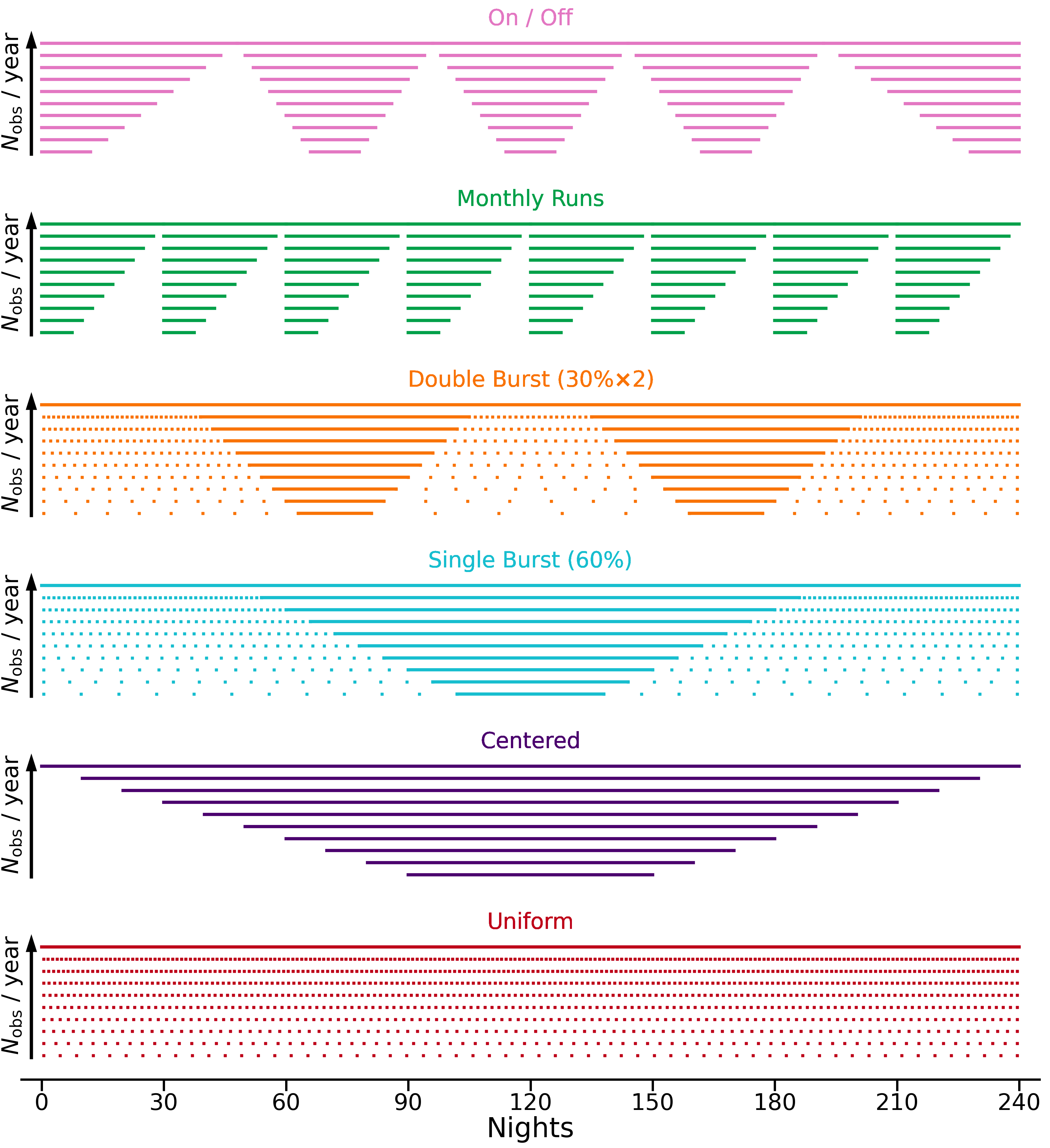}
    \caption{Simulated inter-night observing schedules over a single observing season. We show the idealized distribution of observations across a 240-night baseline for each of the strategies described in Section \ref{sec:internight_sim}. Isolated points represent individual observing nights and lines represent blocks of consecutive observing nights. For each strategy, we show the distribution of observing nights for different numbers of annual observations, increasing in the vertical direction from 60 to 240 in increments of 20 observations per year.}
    \label{fig:fisher_spotstrat}
\end{figure}

Unlike oscillations and granulation, rotationally modulated, activity-induced RV signals exhibit correlations on timescales longer than a single night. This is true for both activity GP kernels shown in \autoref{fig:fisher_covar_long}. To explore how these signals impact long term RV precision, we simulate sets of observations with one visit per night and various inter-night distributions across a typical observing season, which we assume here to be eight months, or 240 days. The following strategies are considered:
\begin{itemize}
    \item \textbf{Uniform Sampling:} Observations are uniformly distributed across each observing season
    \item \textbf{Centered:} All observations occur on consecutive nights at the center of each observing season
    \item \textbf{Single Burst:} A mixture of the uniform and centered strategies, in which 60\% of the observations take place on consecutive nights at the center of each season and the remainder are uniformly distributed
    \item \textbf{Double Burst:} Similar to single burst, but with two sets of consecutive nights bracketing the center of the season, each containing 30\% of the observations
    \item \textbf{Monthly Runs:} Observations occur in equal blocks of consecutive nights at the start of each month
    \item \textbf{On / Off:} Observations occur in five equally spaced blocks of consecutive nights
\end{itemize}
These strategies are depicted in \autoref{fig:fisher_spotstrat}. For each strategy, we build a 10 year observing schedule with annual allocations ranging from 60 observations per year to 240 observations per year, and we use \autoref{eq:fisher} to calculate the expected mass precision for an exoplanet with orbital parameters $K=10$ cm~s$^{-1}$ and $P=300$ d, again marginalizing over orbital phase as in the previous section. We calculate $\sigma_K$ separately for the quasiperiodic and Mat\'ern-5/2 activity GP kernels, and we do not include any other sources of noise in the covariance matrix.

\subsection{Inter-night Observing Strategy Results}\label{sec:internight_res}

\begin{figure} 
    \centering
    \includegraphics[width=1.0\linewidth]{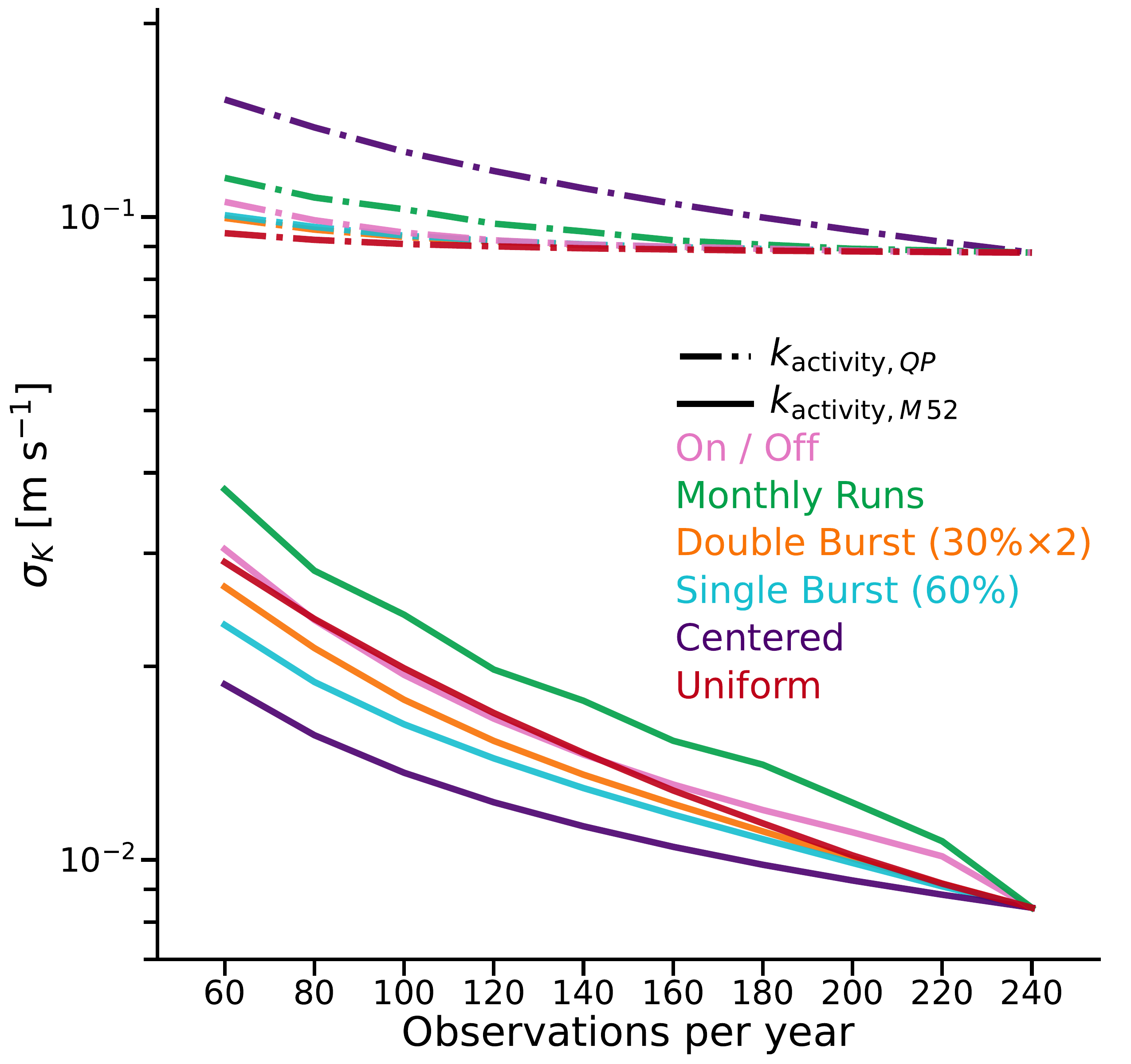}
    \caption{Expected semi-amplitude uncertainty, $\sigma_K$, for a $K=10$ cm~s$^{-1}$ signal and an orbital period of $P=300$ days as a function of number of observations per year. We calculate $\sigma_K$ for simulated observing schedules with 10 years of observations and a 240-day annual observing season, wherein the observations for each season are distributed according to each of the strategies depicted in \autoref{fig:fisher_spotstrat} and described in Section \ref{sec:internight_sim}. We include correlated noise contributions from rotationally-modulated stellar activity represented by a quasiperiodic kernel (dot-dashed lines) and by a Mat\'ern-5/2 kernel (solid lines).}
    \label{fig:fisher_spottests}
\end{figure}

The results from the two activity kernels show a stark contrast in the preferred strategy, the overall magnitude of $\sigma_K$, and the degree to which $\sigma_K$ changes with increasing number of observations (\autoref{fig:fisher_spottests}).
When we build the covariance matrix using the quasiperiodic kernel given in \autoref{eq:qp_act}, the total length of the observing baseline is the overriding factor in determining $\sigma_K$. The Centered strategy, which uses just a fraction of the full 240-day seasonal baseline each year, performs significantly worse than the other strategies. For the other strategies, performance appears to be dictated by the uniformity of the observations across the seasonal baseline; the Uniform strategy produces the smallest uncertainty, followed by the Double Burst, Single Burst, On / Off, and Monthly schedules.
But for the Mat\'ern-5/2 kernel, \autoref{eq:mat_act}, dense coverage is much more important than broad coverage. Strategies with long runs of high-cadence observations (Centered, Single Burst, Two Bursts) consistently produce the smallest $\sigma_K$ for a given number of observations.
And while the performance of all six strategies improves monotonically with increasing number of observations per year, these improvements are much more significant for Mat\'ern-5/2 activity kernel than for the quasiperiodic activity kernel.
The predicted value of $\sigma_K$ decreases by less than $7\%$ for the best performing strategy for the quasiperiodic kernel when the number of observations increases from 60 to 240 per year, while the equivalent change in $\sigma_K$ for the best performing strategy for the Mat\'ern-5/2 kernel is greater then $50\%$.
In addition, we show that although these two kernels nominally represent the covariance behavior of the same physical process, they result in uncertainties that differ by nearly an order of magnitude regardless of observing strategy.

The differences in the predicted uncertainty for each activity kernel can be explained by their respective covariance behaviors.
The Mat\'ern-5/2 kernel's negative covariance on short time scales ($\lesssim10$ days) favors high-cadence observations, which will efficiently average out activity-induced RV variations.
The covariance of the quasiperiodic kernel remains positive on all time scales, so the preferred strategies will be those with separations \rreva{no smaller than the correlation time.
The covariance is close to zero for these strategies, so the measurement uncertainties are in practice equivalent to white noise.}
The local maximum near the stellar rotation period penalizes the Monthly strategy, for which consecutive sets of observations are separated by this same amount, and the relatively poor performance of the high-cadence Centered strategy naturally follows, as clustering all observations in the subset of a season leads to high covariance. 
The time scale for the covariance of the quasi-periodic kernel to fall to zero can also explain the significantly lower $\sigma_K$ values we expect to retrieve for the Mat\'ern-5/2 kernel.
The \rrev{covariance for the Mat\'ern-5/2 kernel and its derivative} not only reaches a white noise approximation at much smaller separations, but the negative covariance allows the activity signal to be averaged out at a rate much faster than white noise binning.

\subsection{Comparison to Posterior Sampling}

\rrev{
To test the accuracy of our methods, we perform a set of injection-recovery tests and compare the resulting parameter uncertainties to the uncertainties we predict from the expected Fisher information content.
We simulate a 10-year RV time series consisting of a planet signal with $\theta_{\rm in}=\{K_{\rm in},P_{\rm in},\phi_{0,\rm in}\}=\{1$ m~s$^{-1}$, $300$ days, $0.6\}$ and a stellar activity signal drawn from a GP seeded by the quasiperiodic kernel given in \autoref{eq:qp_act}. We then generate a set of 1000  measurements randomly sampled across the 10-year baseline, and we introduce white noise by drawing each RV from a Gaussian centered on the true value and with a width of 1 cm~s$^{-1}$. These steps are repeated with different random seeds to produce 10 independent simulated data sets. We also repeat the process for input semi-amplitudes of $K_{\rm in}=10$ cm~s$^{-1}$ and $K_{\rm in}=10$ m~s$^{-1}$ as well as for the Mat\'ern-5/2 activity kernel given in \autoref{eq:mat_act}.
}

\rrev{For each simulated data set, we use the \texttt{numpyro} library \citep{Phan2019} to fit the measured RVs and sample our model parameters, $\theta$, via Markov Chain Monte Carlo (MCMC) analysis. We set Gaussian priors on $K$ and $P$, centered on their input values and with widths of $K_{\rm in}/2$ and $50$ days, respectively, and we sample $\phi_0$ uniformly on [$-\pi,\pi$] with cyclic boundary conditions. The sampler is conditioned on the input GP kernel using \texttt{tinygp} \citep{Foreman-Mackey2024} with the hyperparameters fixed to their input values, and we initialize each fit at $\theta = \{K,P,\phi_0\}=\{K_{\rm in}, 280$ days$, 1.0\}$. The sampler consistently failed to converge for the trials with the quasiperiodic activity kernel and $K_{\rm in}=10$ cm~s$^{-1}$, but we successfully recover the input parameters to better than $2\sigma$ in all other cases. We compute $\sigma_K$ and $\sigma_P$ for each trial as the standard deviations of the posterior parameter distributions for $K$ and $P$.
} 

In \autoref{tab:validation}, we list the recovered parameter uncertainties for each test alongside the expected parameter uncertainties as calculated via Fisher information analysis. The results of the two analyses are fully consistent with each other in all cases for which the sampler successfully recovered the input planet signal, confirming that Fisher information analysis is indeed a valid method for predicting parameter uncertainties. We note that the case for which injection-recovery tests failed to converge ($K_{\rm in}=10$ cm~s$^{-1}$, quasiperiodic kernel), is also the only one for which the expected $\sigma_K$ from the Fisher information calculation is greater than the injected $K$. That is, the expected detection significance is $<1\sigma$, which likely explains the failure to recover the injected signal. \rreva{We test values of $K_{\rm in}$ between 10 cm~s$^{-1}$ and 1 m~s${-1}$ with the quasiperiodic kernel to determine whether this is indeed the case, and we find that our fit begins to converge on the input signal for some simulations with $K_{\rm in}>0.6$ m~s${-1}$ ($K/\sigma_K\approx4$) and it reliably converges by $K_{\rm in}=0.9$ m~s${-1}$ ($K/\sigma_K\approx5$). When the fit converges, the estimated uncertainties are consistent with those predicted by Fisher information analysis.}

\begin{deluxetable*}{lccccc}
\tablecaption{Parameter uncertainties from posterior sampling and Fisher information analysis \label{tab:validation}}
\tablehead{
    \colhead{Activity Kernel} & \colhead{$K_{\rm in}$}&  \multicolumn{2}{c}{$\sigma_K$}&  \multicolumn{2}{c}{$\sigma_P$} \\
    & \multicolumn{1}{c}{(m~s$^{-1}$)} & \multicolumn{2}{c}{(m~s$^{-1}$)} & \multicolumn{2}{c}{(days)}
}
\startdata
    & & MCMC & Fisher & MCMC & Fisher\\
    \hline
    Quasiperiodic & $0.1$ & $0.047\pm0.001$ & $0.179$ & $50.2\pm1.4$ & $24.5$\\
    Quasiperiodic & $1.0$ & $0.173\pm0.005$ & $0.179$ & $2.47\pm0.51$ & $2.44$\\
    Quasiperiodic & $10.0$ & $0.181\pm0.003$ & $0.179$ & $0.232\pm0.005$ & $0.245$\\
    \hline
    Mat\'ern-5/2 & $0.1$ & $0.0221\pm0.0007$ & $0.0235\pm0.0005$ & $3.42\pm0.54$ & $3.27\pm0.12$\\
    Mat\'ern-5/2 & $1.0$ & $0.0237\pm0.0005$ & $0.0235\pm0.0004$ & $0.319\pm0.008$ & $0.327\pm0.009$\\
    Mat\'ern-5/2 & $10.0$ & $0.0241\pm0.0004$ & $0.0235\pm0.0006$ & $0.032\pm0.001$ & $0.032\pm0.001$\\
\enddata
\tablenotetext{}{Values are given as the median $\pm$ standard deviation across the 10 trials. We do not list the standard deviation of the Fisher information results for the quasiperiodic activity kernel, as the spread in both $\sigma_K$ and $\sigma_P$ is negligible ($<1$ part per thousand) for all values of $K_{\rm in}$.}
\end{deluxetable*}

\section{Application to a realistic survey}\label{sec:fisher_simulations}

Our analysis in the previous section sheds some light on the relative performance of different observing cadences both within a night and across many nights. However, the implications of the \textit{absolute} precision we report, quantified by the predicted uncertainty on $K$, are not as easy to interpret given that we have isolated individual noise sources. That is, these results alone do not tell us how each \rrev{strategy} would fare in the presence of all sources of correlated noise. In addition, we ignored practical constraints on our ability to reproduce each intended survey strategy. Real observations are subject to constraints from telescope and observatory schedules, weather losses, nightly and seasonal observing windows, and intra-survey competition.
In this section, we explore how our choice of survey strategy affects sensitivity to Earth analogs in the presence of all sources of correlated stellar signals described in Section \ref{sec:fisher_covariance}. Accounting for realistic observing constraints, we apply several of the strategies assessed in Section \ref{sec:unit_tests} to simulations of a full EPRV exoplanet survey and calculate the resulting Fisher information content.

\subsection{Survey Parameters and Schedule Simulation}

As a template for our survey simulations, we adopt the target list of the NEID Earth Twin Survey \citep[NETS;][]{Gupta2021} and the parameters and constraints of the Terra Hunting Experiment \citep{Hall2018}. The Terra Hunting Experiment will nominally have access to 50\% of the total observing time on the HARPS-3 spectrograph \citep{Thompson2016} for 10 years. We compute the total time expected to be available for observations and allocate this evenly across all targets. We consider both a 40-star target list as well as a narrower, 20-star subset of this list, so that we may evaluate the effects of sample size on survey performance.  Targets are set to be observed 100 times per year for the former sample and 200 times per year for the latter sample. The targets for both lists are given in \autoref{tab:fisher_targets}. Following our findings in Section \ref{sec:intranight_res}, just one visit is allocated on each night that a star is observed.
See Appendix for a detailed description of the target selection criteria and time allocation breakdown.

Because of the wide variation in our results in Section \ref{sec:internight_res}, we consider four different inter-night observing strategies for our survey simulations. These include the best- and worst-case strategies for the quasiperiodic and Mat\'ern-5/2 activity kernels (Uniform, Centered, and Monthly) as well as the Double Burst strategy as an intermediate reference point.
We simulate a 10-year survey for each of these strategies using a custom scheduling framework. In brief, our scheduler accounts for the following practical observing constraints: target observability, seeing, weather losses, telescope and facility maintenance, and intra-survey competition (i.e., we can not observe two or more targets simultaneously). Quantitative descriptions of the scheduling algorithms are given in Appendix .

\begin{deluxetable}{lccccc}
\tablecaption{Target Sample for EPRV Survey Simulations \label{tab:fisher_targets}}
\tablehead{
    \colhead{Star Name}&  \colhead{$V^\dagger$}&  \colhead{$T_{\rm eff}^\dagger$}&  \colhead{$\log g^\dagger$} & \colhead{$M_\star^\dagger$} & \colhead{Season} \\
    \colhead{}&  \colhead{(mag)}&  \colhead{(K)}&  \colhead{(log (cm s$^{-2}$))} & \colhead{($M_\odot$)} & \colhead{(days)}
}
\startdata
HD4614* & $3.46$ & 5919 & $4.37$ & $0.99$ & 284 \\
HD10700 & $3.49$ & 5333 & $4.60$ & $0.78$ & 211 \\
HD19373* & $4.05$ & 5938 & $4.18$ & $1.18$ & 275 \\
HD26965 & $4.43$ & 5092 & $4.51$ & $0.79$ & 229 \\
HD185144* & $4.67$ & 5242 & $4.56$ & $0.81$ & 331 \\
HD34411* & $4.70$ & 5873 & $4.26$ & $1.12$ & 271 \\
HD115617 & $4.74$ & 5562 & $4.44$ & $0.94$ & 210 \\
HD182572 & $5.17$ & 5587 & $4.15$ & $1.11$ & 287 \\
HD201091* & $5.21$ & 4361 & $4.63$ & $0.68$ & 298 \\
HD10476* & $5.24$ & 5190 & $4.51$ & $0.84$ & 260 \\
HD157214* & $5.38$ & 5817 & $4.61$ & $0.96$ & 313 \\
HD86728* & $5.40$ & 5742 & $4.31$ & $1.09$ & 268 \\
HD143761* & $5.40$ & 5833 & $4.29$ & $1.00$ & 305 \\
HD146233 & $5.50$ & 5785 & $4.41$ & $1.04$ & 261 \\
HD55575 & $5.55$ & 5888 & $4.32$ & $0.97$ & 272 \\
HD10780* & $5.63$ & 5344 & $4.54$ & $0.90$ & 282 \\
HD190360* & $5.70$ & 5549 & $4.29$ & $1.00$ & 301 \\
HD4628 & $5.74$ & 4937 & $4.54$ & $0.75$ & 244 \\
HD50692 & $5.76$ & 5913 & $4.39$ & $1.02$ & 266 \\
HD172051 & $5.85$ & 5636 & $4.58$ & $0.91$ & 224 \\
HD52711 & $5.93$ & 5886 & $4.39$ & $1.03$ & 267 \\
HD110897* & $5.95$ & 5911 & $4.49$ & $0.90$ & 294 \\
HD186408 & $5.96$ & 5778 & $4.28$ & $1.08$ & 315 \\
HD38858 & $5.97$ & 5735 & $4.46$ & $0.95$ & 236 \\
HD187923* & $6.16$ & 5774 & $4.23$ & $1.03$ & 285 \\
HD217107 & $6.18$ & 5575 & $4.25$ & $1.04$ & 241 \\
HD186427 & $6.20$ & 5747 & $4.37$ & $1.04$ & 315 \\
HD126053 & $6.25$ & 5714 & $4.54$ & $0.91$ & 263 \\
HD168009* & $6.30$ & 5808 & $4.33$ & $1.05$ & 321 \\
HD127334* & $6.36$ & 5671 & $4.27$ & $1.07$ & 308 \\
HD166620 & $6.38$ & 4970 & $4.51$ & $0.78$ & 317 \\
HD9407* & $6.52$ & 5672 & $4.45$ & $1.00$ & 286 \\
HD179957* & $6.75$ & 5741 & $4.42$ & $1.08$ & 322 \\
HD221354* & $6.76$ & 5221 & $4.47$ & $0.86$ & 291 \\
HD154345* & $6.76$ & 5455 & $4.52$ & $0.89$ & 322 \\
HD68017 & $6.78$ & 5626 & $4.60$ & $0.86$ & 263 \\
HD24496 & $6.81$ & 5531 & $4.50$ & $0.94$ & 254 \\
HD170657 & $6.81$ & 5040 & $4.54$ & $0.78$ & 239 \\
HD51419* & $6.94$ & 5732 & $4.51$ & $0.90$ & 265 \\
HD116442 & $7.06$ & 5155 & $4.54$ & $0.75$ & 254 \\
\enddata
\tablenotetext{*}{Included in 20-star subsample}
\tablenotetext{{\dagger}}{Parameters taken from \citet{Brewer2016}.}
\end{deluxetable}

\subsection{RV Noise Contributions}

As in Section \ref{sec:unit_tests}, we do not need to generate RV measurements to accompany our simulated survey schedules. The Fisher information content does not depend on the measurements themselves. However, we do need to account for the associated noise properties. Here, we describe the computation of the covariance matrix for each target star, accounting for all four sources of noise in \autoref{eq:covar}.

\subsubsection{Photon Noise}\label{sec:sim_covar_phot}

Photon noise contributes a diagonal term to each stars covariance matrix.
The photon noise precision of an exposure, $\sigma_{\rm photon}$, depends on the RV information content of the observed spectrum as described by \citet{Bouchy2001}. For observations with a given spectrograph, the precision will vary as a function of the spectral type of the star and the achieved SNR, which in turn depends on the exposure time, stellar brightness, and observing conditions.
To determine $\sigma_{\rm photon}$ for our simulated observations, we first calculate the expected SNR as a function of $T_{\rm eff}$, $V$-band magnitude, and exposure time using the NEID exposure time calculator. We then apply two multiplicative corrections to account for the simulated seeing conditions and for the effective system throughput of HARPS-3 relative to NEID
\begin{equation}\label{eq:snrscale}
    {\rm SNR}_{\rm scaled} = {\rm SNR}\sqrt{R_{\rm seeing}R_{\rm system}}.
\end{equation}
The seeing correction is calculated by taking the cross section of the fiber \citep[on-sky diameter $d_f=1.4''$;][]{Thompson2016} and the seeing disk, assuming that the star is perfectly centered
\begin{equation}
    R_{\rm seeing} = \frac{1-e^{-\frac{1}{2}\left(\frac{d_f/2}{{\rm FWHM}/2.355}\right)^2}}{1-e^{-\frac{1}{2}\left(\frac{d_f/2}{{\rm FWHM}_{\rm median}/2.355}\right)^2}},
\end{equation}
and comparing this to the same cross section in median seeing conditions (${\rm FWHM}_{\rm median}=1.3''$).
To scale the throughput, we calculate the photon noise precision precision we would expect to achieve for some test values of $R_{\rm system}$, and find we can match the HARPS-3 results calculated by \citet{Thompson2016} if we apply a correction factor of $R_{\rm system} = 1.375$. However, \citet{Thompson2016} assume a fairly conservative system throughput of $5\%$. Here, we assume a more optimistic $10\%$ average efficiency, so we set $R_{\rm system} = 2.75$. The final photon noise is then estimated from ${\rm SNR}_{\rm scaled}$ using the NEID exposure time calculator as described in \citet{Gupta2021}. We note that for NEID, the detector saturates at SNR$\approx625$. In median seeing conditions, we find that ${\rm SNR}_{\rm scaled}$ will exceed this threshold in a 320 second exposure for stars brighter than $V=4.68$, hence our decision to use shorter exposures for the brightest stars in our sample.
\rreva{We do not account for the dependence of throughput on target airmass.}

\subsubsection{Stellar Variability}
We calculate the correlated noise contributions from stellar variability using Equations \ref{eq:kosc} and \ref{eq:kgran} for oscillations and granulation, respectively, again integrating over the exposure time as in Section \ref{sec:intranight_sim}, and using Equations \ref{eq:qp_act} and \ref{eq:mat_act} for spot-induced activity.
In Section \ref{sec:unit_tests}, we assumed perfectly \rreva{known} solar hyperparameters as given in \autoref{tab:fisher_gp_hyper}. We continue to assume Solar values for the activity kernels here, but to enable a more even comparison of their absolute performance, we set the amplitude of the quasiperiodic kernel to be $\alpha=0.6$ m~s$^{-1}$ such that it matches that of the Mat\'ern-5/2 kernel at $\Delta=0$. For the asteroseismic signals, we take advantage of known stellar parameter scaling relations \citep[given in ][]{Luhn2023} to estimate more accurate oscillation and granulation kernel hyperparameters for each of our target stars. As inputs to these scaling relations, we take spectroscopically derived effective temperatures, $T_{\rm eff}$, and surface gravities, $\log g$, from \citet{Brewer2016}. These stellar parameters are listed in \autoref{tab:fisher_targets}.
For each simulated survey schedule, we treat the two activity kernels independently just as we did in Section \ref{sec:unit_tests}, computing separate covariance matrices and analyzing separate sets of results.

\section{Results}\label{sec:fisher_results}

\subsection{Detection Sensitivity and Survey Success Metrics}\label{sec:fisher_success_metrics}
\begin{figure*}
  \centering
  \includegraphics[width=\textwidth]{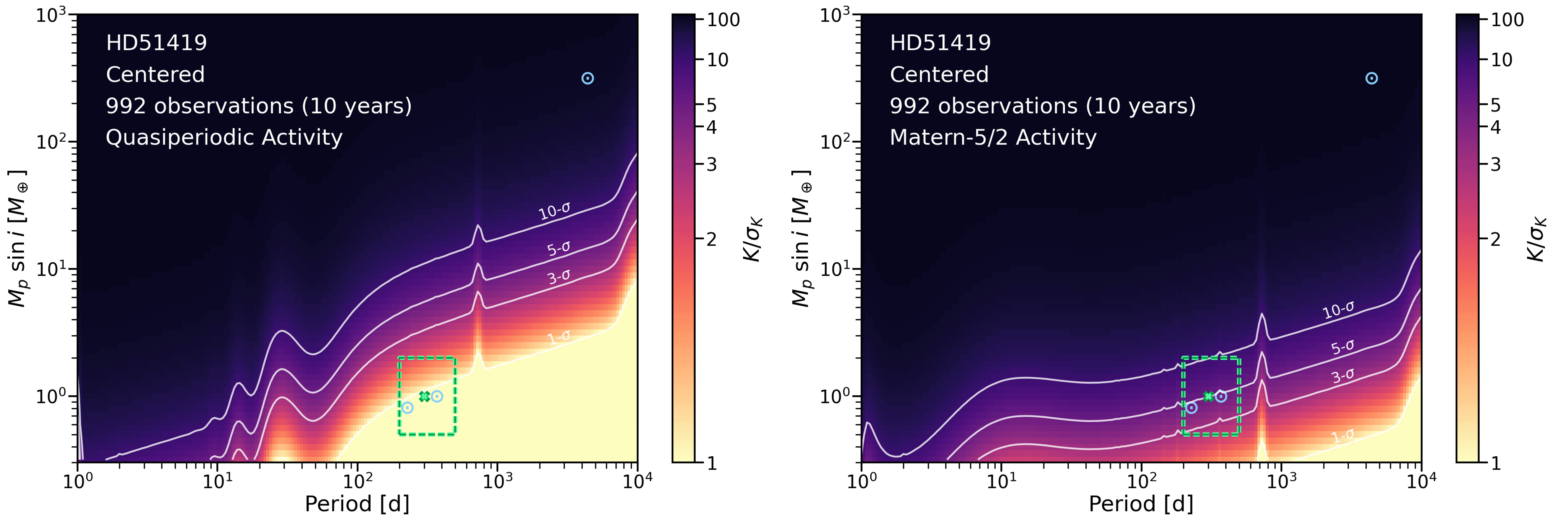}
\caption{Expected exoplanet detection sensitivity limits as calculated using our Fisher information framework for a typical target star with approximately 100 observations per year for 10 years and the Centered inter-night observing strategy. We include contributions from photon noise, oscillations, granulation, and stellar activity. We plot $K/\sigma_K$ as a function of minimum mass, $M_p \sin i$, and orbital period, $P$.  For reference, we also plot the 3-, 5-, and 10-$\sigma$ detection significance contours and we indicate the locations of the Solar System planets Venus, Earth, and Jupiter with the $\odot$ symbol. The green `x' and green dashed box indicate the mass and period values and ranges we use to define an Earth analog for the $[K/\sigma_K]_\oplus$ and $D_\oplus$ survey success metrics, respectively.
Results for the quasiperiodic activity kernel are shown on the left and results for the Mat\'ern-5/2 activity kernel are shown on the right.}
\label{fig:fisher_sensitivity}
\end{figure*}

With the simulated schedules and covariance matrices in hand, we can use Equations \ref{eq:fisher} and \ref{eq:fisher_sig} to calculate $\sigma_K$ and assess the achieved sensitivity of each survey strategy.
We first compute $K$ as in \autoref{eq:semiamp} for a broad range of $M_p \sin i - P$ parameter combinations, assuming circular orbits and taking stellar masses \rrev{listed in \autoref{tab:fisher_targets} \citep[taken from][]{Brewer2016}},  and then we calculate $\sigma_K$ for the corresponding parameter vector $\theta$, marginalizing over orbital phase at each point. We use these results to determine the expected detection significance, $K/\sigma_K$, which we show for two representative cases in \autoref{fig:fisher_sensitivity}.

To compare performance across different survey strategies, we define two quantitative success metrics.
The first metric is simply the detection significance for an Earth analog, $[K/\sigma_K]_\oplus$. Here, we define an Earth analog to be a $M_p \sin i = M_\oplus$ exoplanet with an orbital period of 300 days (green `x' in \autoref{fig:fisher_sensitivity}). We choose $P=300$ days rather than $P=365$ days so as not to bias our results against strategies that are more susceptible to annual aliasing. For each simulated survey, we calculate and tabulate (\autoref{tab:fisher_results}) the number of stars for which the detection significance exceeds a threshold $H$, i.e., $N_{[K/\sigma_K]_\oplus>H}$, setting $H=[3,5,10]$ for the Mat\'ern-5/2 activity kernel results and $H=1$ for the quasiperiodic kernel results. We also show the distribution of $[K/\sigma_K]_\oplus$ values for each simulated survey in \autoref{fig:fisher_progression}.

\begin{deluxetable*}{lcrrrrrrrr}
\tablecaption{Survey Simulation Success Metrics \label{tab:fisher_results}}
\tablehead{
    \colhead{Activity Kernel} & \colhead{Metric} & \colhead{Centered}&  \colhead{Centered}&  \colhead{Uniform}&  \colhead{Uniform} & \colhead{Monthly}&  \colhead{Monthly}&  \colhead{Double Burst}&  \colhead{Double Burst}\\
    \colhead{}&  \colhead{}&  \colhead{(40 stars)}&  \colhead{(20 stars)} &  \colhead{(40 stars)}&  \colhead{(20 stars)} &  \colhead{(40 stars)}&  \colhead{(20 stars)} &  \colhead{(40 stars)}&  \colhead{(20 stars)}
}
\startdata
Quasiperiodic & $N_{[K/\sigma_K]_\oplus>1}$ & 5 & 19 & 38 & 20 & 25 & 19 & 38 & 20 \\
Mat\'ern-5/2 & $N_{[K/\sigma_K]_\oplus>3}$ & 40 & 20 & 40 & 20 & 38 & 20 & 40 & 20 \\
Mat\'ern-5/2 & $N_{[K/\sigma_K]_\oplus>5}$ & 21 & 19 & 22 & 19 & 4 & 18 & 17 & 19 \\
Mat\'ern-5/2 & $N_{[K/\sigma_K]_\oplus>10}$ & 0 & 2 & 0 & 1 & 0 & 1 & 0 & 2 \\
\hline
Quasiperiodic & $\overline{D_\oplus} (H=1)$ & 0.40 & 0.28 & 0.59 & 0.30 & 0.52 & 0.29 & 0.58 & 0.30\\
Mat'ern-5/2 & $\overline{D_\oplus} (H=3)$ & 0.85 & 0.49 & 0.84 & 0.48 & 0.68 & 0.48 & 0.80 & 0.49\\
Mat'ern-5/2 & $\overline{D_\oplus} (H=5)$ & 0.49 & 0.37 & 0.48 & 0.36 & 0.32 & 0.35 & 0.44 & 0.37\\
Mat'ern-5/2 & $\overline{D_\oplus} (H=10)$ & 0.07 & 0.13 & 0.06 & 0.12 & 0.01 & 0.11 & 0.04 & 0.12\\
\enddata
\tablenotetext{}{\rrev{The given values represent the number of stars, $N$, and the average fractional parameter space, $\overline{D_\oplus}$, above each specified detection significance threshold.}}
\end{deluxetable*}

We adopt a slightly looser definition of an Earth analog for our second success metric, $D_\oplus$, which we will refer to \rrev{as the fractional} Earth analog discovery space:
\begin{equation}\label{eq:terraspace}
    D_\oplus = \frac{\int_{P_{\rm min}}^{P_{\rm max}}\log\left(\frac{2M_\oplus}{M_{p,{\rm min}}}\right)d\log P}{\int_{200 {\rm d}}^{500 {\rm d}}\log\left(\frac{2M_\oplus}{0.5M_\oplus}\right)d\log P}.
\end{equation}
We set $P_{\rm min}=200$ days and $P_{\rm max}=500$ days and we calculate the minimum detectable mass, $ M_{p,{\rm min}}$, as
\begin{equation}
    M_{p,{\rm min}} =
    \begin{cases}
          0.5M_\oplus & [M_p \sin i]_{H} < 0.5M_\oplus\\
          [M_p \sin i]_{H} & [M_p \sin i]_{H}\geq 0.5M_\oplus\\
    \end{cases}
\end{equation}
where $[M_p \sin i]_{H}$ is the mass at which $K/\sigma_K=H$ for a given orbital period. \rrev{The value of $D_\oplus$ thus represents the fraction of parameter space to which each survey is sensitive in the region bounded by$200$ days $<P<$ 500 days and $0.5M_\oplus <M_p \sin i < 2 M_\oplus$.}
We again set thresholds of $H=3$, 5, and 10 for the Mat\'ern-5/2 activity kernel and a $H=1$ threshold for the quasiperiodic activity kernel. By expanding our definition of an Earth analog to include all exoplanets \rrev{in this region, indicated} by the green box in \autoref{fig:fisher_sensitivity}, we reduce our sensitivity to small $K/\sigma_K$ fluctuations in $M_p \sin i - P$ space.
For each simulated survey, we \rrev{take the mean value of} $D_\oplus$ for all stars and show the results in \autoref{tab:fisher_results}. \rrev{We divide this metric in half for the 20-star sample results to account for the smaller survey sample.}

\subsection{Survey Strategy Performance Comparison}

\subsubsection{Quasiperiodic activity model}

\begin{figure*}
  \centering
  \includegraphics[width=\textwidth]{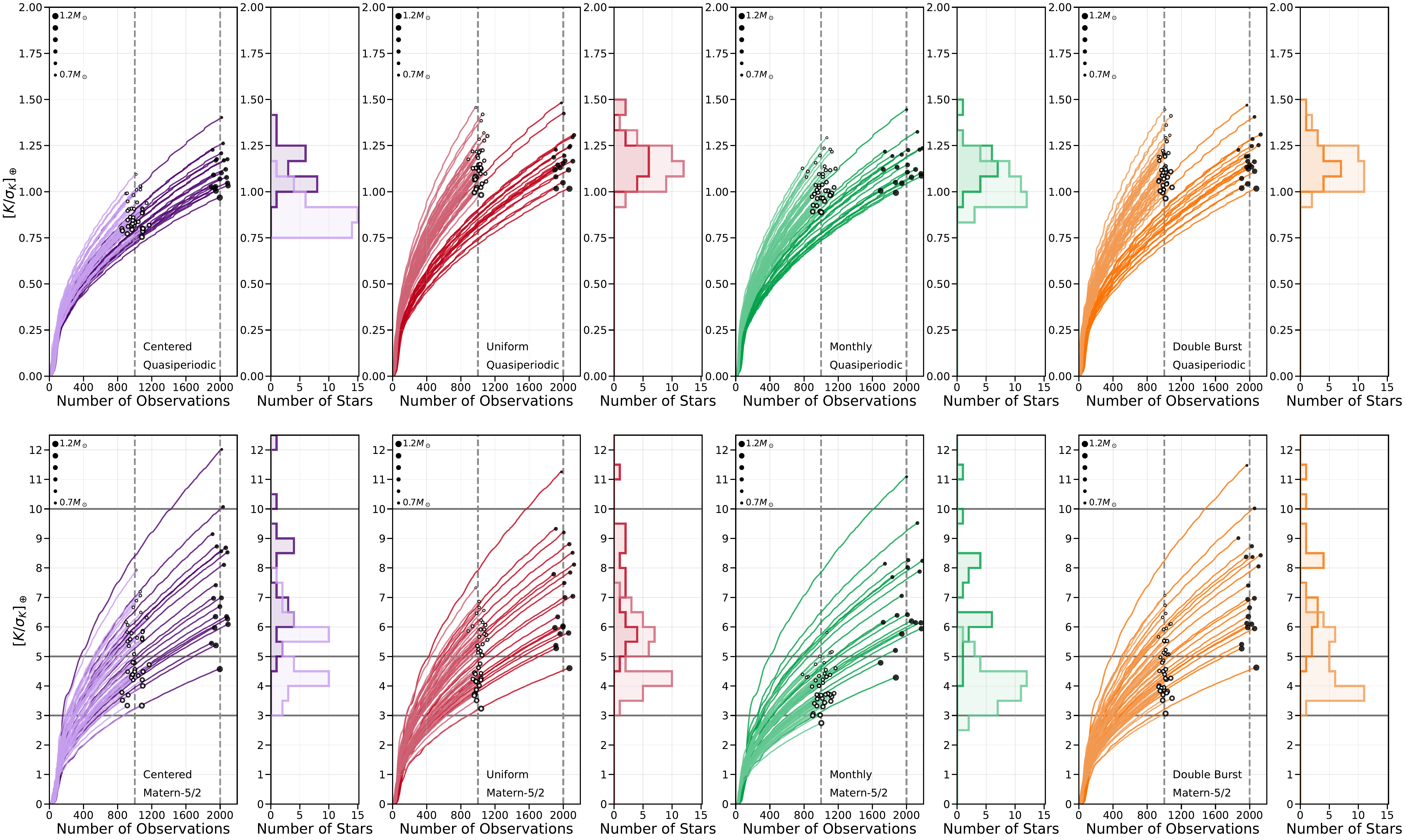}
\caption{Expected Earth analog detection significance, $[K/\sigma_K]_\oplus$, as a function of number of visits for each of our survey simulations. The darker lines that terminate in filled circles correspond to the results for the 20-star simulations and the lighter lines that terminate in open circles correspond to the results for the 40-star simulations. The size of each circle scales with stellar mass. We note that each of these lines is exhibits a relatively smooth dependence on number of observations within an observing season, but that these smooth changes are interrupted by sharper kinks between observing seasons.  Dashed vertical lines mark the expected number of observations per star for an idealized observing schedule, i.e., 1000 total observations for each of the 40-star simulations and 2000 total observations for each of the 20-star simulations. The final distributions of $[K/\sigma_K]_\oplus$ values for each survey are shown as histograms in the narrow panels to the right side.}
\label{fig:fisher_progression}
\end{figure*}

In the case that our stellar activity model is best represented by the quasiperiodic kernel given in \autoref{eq:qp_act}, \rrev{all} of the survey strategies tested here are expected to yield \rrev{marginal} detections of Earth analogs\rrev{, at best} (\rrev{$[K/\sigma_K]_\oplus<1.5$;} \autoref{fig:fisher_progression}). \rrev{Still,} we \rrev{can assess} the relative performance of the different strategies. As we show in \autoref{tab:fisher_results} and \autoref{fig:fisher_progression}, the $N_{[K/\sigma_K]_\oplus>1}$ success metric is fairly sensitive to our choice of survey strategy for the full 40-star sample. While nearly every star reaches the $[K/\sigma_K]_\oplus>1$ threshold for the Uniform and Double Burst strategies, far fewer stars reach this threshold for the Monthly strategy and only a handful do for the Centered strategy. Yet in the case of the 20-star subsample, this performance gap is nearly erased, and nearly every star reaches the $[K/\sigma_K]_\oplus>1$ threshold for all strategies. This is entirely consistent with our interpretation of the \rrev{quasiperiodic kernel analysis} in Section \ref{sec:internight_res}. For this activity kernel, the achieved detection sensitivity is strongly correlated with the length of the observing baseline within each observing season. The increased number of observations only significantly benefits the Centered strategy, for which this increase also translates to a greatly extended annual baseline \rrev{(\autoref{fig:fisher_progression})}. \rrev{There is a nearly} negligible change in the distribution of $[K/\sigma_K]_\oplus$ values between the 40-star and 20-star samples for the Uniform and Double Burst strategies, despite the latter sample having twice the number of observations per star.

\rrev{The $D_\oplus$ success metrics paint a more granular picture of relative survey performance than do the $[K/\sigma_K]_\oplus$ values, but the results of these two metrics in \autoref{tab:fisher_results} are largely consistent with each other. The only stark discrepancy is in the results of the Centered strategy with the 40-star sample, for which only the $D_\oplus$ metric is comparable to the other strategies and sample sizes but far fewer stars reach the $[K/\sigma_K]_\oplus>1$ threshold. This suggests that many stars fall just short of the threshold in this case, which is confirmed in \autoref{fig:fisher_progression}.}

\subsubsection{Mat\'ern-5/2 activity model}

For the Mat\'ern-5/2 activity kernel, the absolute performance of each survey strategy is \rrev{much greater than for the quasiperiodic kernel}\rrev{. N}early every star reaches the $H=3$ threshold for all simulated survey strategies and even the $H=5$ threshold for each of the 20-star simulations. And while only half of the observed stars reach $[K/\sigma_K]_\oplus>5$ for the 40-star simulations with the  Centered, Uniform, and Double Burst strategies, each of these strategies will still yield close to 20 stars that are sensitive to Earth analogs at this level. The relatively poor performance of the Monthly strategy here is consistent with our results in Section \ref{sec:internight_res}, as is the lack of variation from strategy to strategy for the 20-star sample. But \rrev{contrary to what one might expect based on the earlier unit tests,} the Centered strategy fails to outpace the Uniform and Double Burst strategies for the 40-star sample. The difference between the results presented here and those presented in Section \ref{sec:internight_res}, which uses an idealized observing schedule, likely stems from the practical constraints imposed on the full survey simulations. By restricting the available observing nights to account for these constraints, we have effectively extended the annual observing baseline of the Centered strategy such that it is no longer as distinct. 

The larger sample size of 40 stars will be preferred for a survey with a target detection threshold of $H=3$. But as $H$ increases to $H=5$ and again to $H=10$, $N_{[K/\sigma_K]_\oplus>H}$ drops off much more steeply for the 40-star sample than for the 20-star sample. This same effect is illustrated in \autoref{fig:fisher_progression}, which shows that the average detection significance is significantly higher for the \rrev{2}0-star sample, thus demonstrating that preferred sample size is inversely proportional to the stringency of our detection significance requirements. This is in stark contrast to our interpretation of the $N_{[K/\sigma_K]_\oplus>H}$ success metric for the quasiperiodic kernel, for which we see only a marginal increase in detection significance when we reduce the sample size and double the number of observations per target.

\rrev{As with the quasiperiodic activity model, the $D_\oplus$ success metric is more granular than -- but largely consistent with -- the $[K/\sigma_K]_\oplus$ metric. There is again one instance in which the performance of a single strategy appears to perform poorly based on the number of stars above a $[K/\sigma_K]_\oplus$ threshold (in this case, the Monthly strategy with the 40-star sample), but the $D_\oplus$ values are more comparable to those of the other strategies.}

\subsection{Cadence vs. Baseline}\label{sec:fisher_cadence_baseline}

The factors driving the relative performance of the survey strategies explored here can largely be distilled into to three key parameters. These are the frequency with which each star is observed, the total number of stars in the sample, and the observing baseline for each star both within each season and across the entire survey. Within the constraints of a survey with a fixed total time allocation, these parameters cannot be independently tuned. But we can still examine how our results change when two or more parameters are adjusted.
Indeed, we have already shown that\rrev{, concerning the trade-off between sample size and cadence, the} results for the two activity kernels are at odds with each other. For the quasiperiodic kernel, increasing the cadence adds little value and is not worth the decrease in sample size, while for the Mat\'ern-5/2 kernel, a higher cadence and smaller sample size is increasingly preferred as we raise our target detection significance threshold. 

\begin{figure*}
  \centering
  \includegraphics[width=\textwidth]{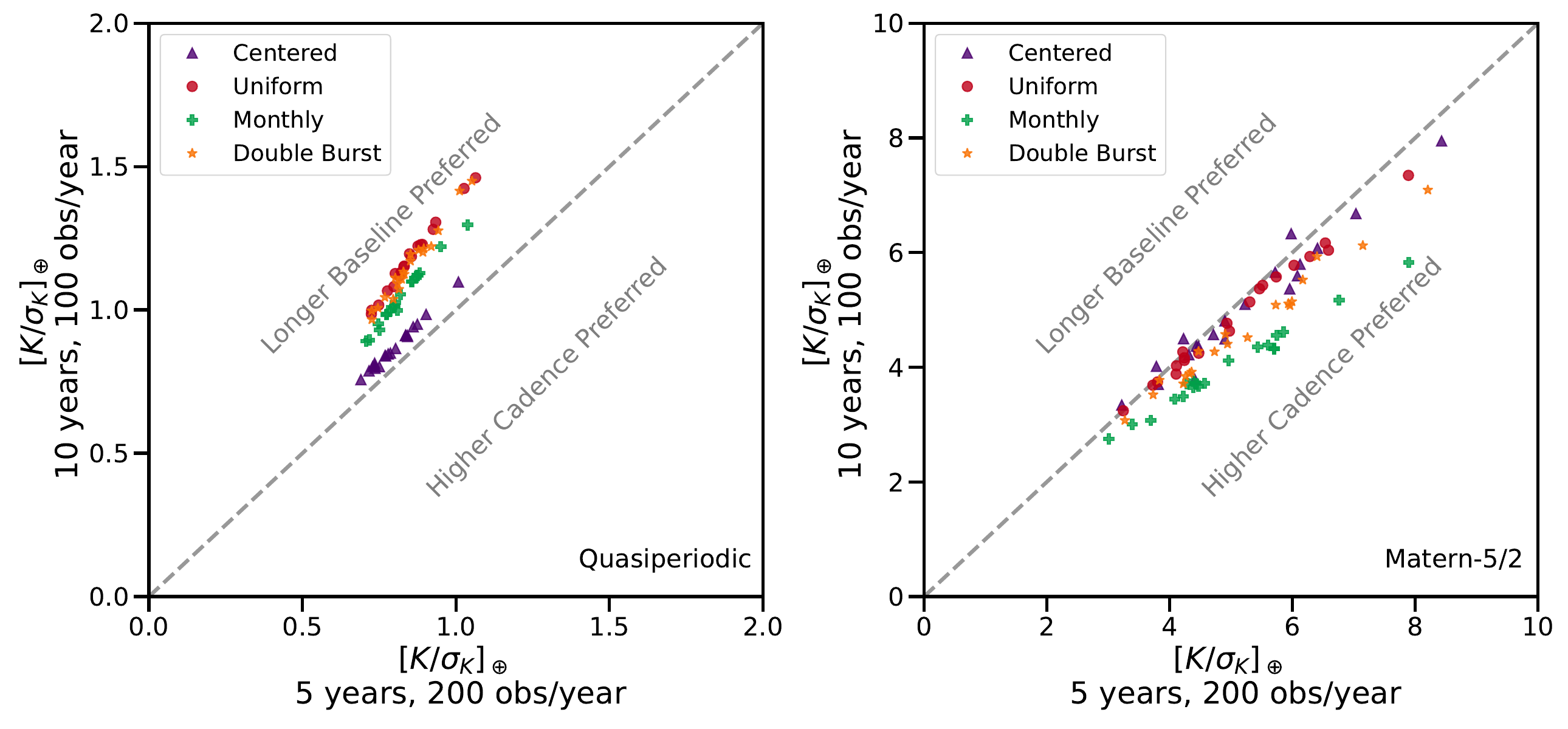}
\caption{Comparison of $[K/\sigma_K]_\oplus$ for a 10-year survey with 100 observations per year (vertical axis) and a 5-year survey with 200 observations per year (horizontal axis). We compute the latter values using the first 5 years of observations from the 20-star survey simulations. On the left, we show that for the quasiperiodic kernel, the longer 10-year survey delivers a higher detection significance than the higher cadence 5-year survey. On the right, we show that the opposite is true for the Mat\'ern-5/2 kernel, which performs better with a higher observing cadence.}
\label{fig:fisher_baseline_cadence}
\end{figure*}

To shed light on the trade-off between observing cadence and total survey baseline, we re-compute the Fisher information and resulting $[K/\sigma_K]_\oplus$ values for each of the 20-star survey simulations using just the first 5 years of observations for each star.  By fixing the total number of observations to about 1000 per star, we can directly compare these results the results for these same stars over the full 10-year duration of the 40-star survey simulations. As \autoref{fig:fisher_baseline_cadence} shows, our strategy recommendation will again depend on the assumed activity kernel. For the quasiperiodic model, the longer 10-year survey outperforms the shorter 5-year survey for all strategies, and for the Mat\'ern-5/2 kernel the shorter, higher-cadence survey wins out. In both cases, the Centered observing strategy is shown to be least sensitive to this trade-off between baseline and cadence. This is consistent with expectations given that changing the number of observations for this strategy serves only to adjust the length of the baseline within each season rather than the frequency of observations.

Based on our results, the preferred survey strategies differ significantly for each of the two activity kernels we have applied, as does the absolute detection significance we can expect to achieve.
But when interpreting these results, one should be careful not to attribute the differences purely to the kernel forms. As we show in \autoref{fig:fisher_covar_long} and \autoref{tab:fisher_gp_hyper}, we assume significantly different correlation length hyperparameters for the quasiperiodic and Mat\'ern-5/2 kernels. While the form of each kernel undoubtedly plays a role in the resulting detection sensitivities, the stark contrast we predict could likely be reduced by adjusting these hyperparameters. We comment on this further in the following section.

\section{Discussion}\label{sec:fisher_discussion}

\subsection{Stellar Variability Model Assumptions}\label{sec:fisher_gpmodel_assumptions}

One shortcoming of the work presented herein is that our analysis framework \rrev{relies on} perfect knowledge of the stellar variability kernels and hyperparameters with which we construct the covariance matrices for our Fisher information calculations.
The target lists for ongoing EPRV exoplanet searches are selectively composed of intrinsically ``quiet'' stars \citep{Brewer2020,Gupta2021}, as is the target list used for the survey simulations in this work. This is advantageous in that it naturally reduces the impact of variability on exoplanet detection, but it has also limited our ability to directly measure and model these stars’ intrinsic variability signals, as the amplitudes of these signals are at a level below the precision floor achieved by surveys with older spectrographs. As such, the stellar variability models we use here are informed primarily by Solar observations and models and, in the case of oscillations and granulation, extended to other stars via scaling relations. While physically motivated, these scaling relations are not perfectly accurate, and oscillation and granulation timescales and amplitudes will often depend on additional stellar physics that these relations do not capture \citep{Cattaneo2003,Jimenez-Reyes2003,Bonanno2014,Garcia2014,Mathur2019,Gupta2022}. The results of our stellar activity kernel analysis should be treated with caution as well\rrev{; the achieved survey sensitivity and our interpretation of the preferred strategy will depend on the kernel form and hyperparameters and differs greatly between the two kernels we test. Empirical investigations of the activity signals of quiet stars, such as the collaborative EXPRES Stellar Signals Project \citep{Zhao2020,Zhao2022} will enable informed observing schemes that are tailored to individual stars.}

We also note that we have assumed that each of the stellar variability kernel hyperparameters are not only perfectly accurate, but also perfectly precise. Even if one were to apply this analysis to stars with well constrained asteroseismic and activity signals, the resulting orbital parameter uncertainties would not include propagated uncertainties on the measured correlated noise hyperparameters. Extending this framework to accommodate hyperparameter uncertainties once empirical models are in hand will improve the accuracy of our results and enable us to quantify how tightly the kernel hyperparameters need to be constrained to adequately model out stellar variability signals and enable Earth analog exoplanet detection.

\subsection{Accounting for Instrumental Noise}\label{sec:fisher_instruments}

We have chosen to ignore noise contributions from spectrograph and pipeline systematics in this study so that we could isolate the effects of other sources of noise and assess relevant mitigation strategies. As we state in Section \ref{sec:fisher_covariance}, we assumed to first order that well-understood instrumental variations are calibrated out prior to computing the RV time series from observed stellar spectra, and we ignore residuals and other poorly understood systematic signals that remain. A thorough discussion of the cumulative covariance contributions from numerous sub-m s$^{-1}$ instrument systematics is well beyond the scope of this work, but it is worth considering their impact on our conclusions.

We repeat the Fisher information analysis for our intra-night observing schedule simulations in Section \ref{sec:unit_tests} with the addition of a single instrumental covariance contribution that is constant for all observations taken within a single night and perfectly uncorrelated across multiple nights. Noise such as this might stem from variations in a wavelength solution that is anchored to a nightly calibration sequence. 
This covariance contribution can be represented by a simple step function kernel:
\begin{equation}
    k_{\rm instrument}(\Delta) = \begin{cases} 
      \sigma_{\rm instrument}^2 & \Delta \leq 0.5 {\rm\ days}\\
      0 & \Delta > 0.5 {\rm\ days}.
   \end{cases}
\end{equation}
We perform this analysis for internal single epoch precisions of $\sigma_{\rm instrument}=30$ cm~s$^{-1}$ and $\sigma_{\rm instrument}=10$ cm~s$^{-1}$, where the former is based on recent best estimates for current generation EPRV spectrographs \citep{Halverson2016,Blackman2020} and the latter is an optimistic projection for future instruments.
As expected, the addition of $k_{\rm instrument}$ to our covariance matrix neither changes the preferred intra-night observing strategy nor accentuates any differences between strategies. Because the covariance timescale is set such that each simulated schedule receives the same number of independent epochs, this instrumental noise contribution serves only to raise the achieved precision floor for all strategies. We find that the increase in $\sigma_K$ can be approximated as
\begin{equation}
    \sigma_{K}^2 \approx \sigma_K^2 + 2\sigma_{\rm instrument}^2/N_{\rm nights}.
\end{equation}
This will have a negligible effect on sensitivity to Earth analogs when $\sigma_K$ is on the order of $10$ cm~s$^{-1}$ or greater (e.g., for our full simulation results using the quasiperiodic activity kernel). But even with thousands of observations and an instrumental precision of $\lesssim30$ cm~s$^{-1}$, this can significantly degrade the achieved Earth analog detection significance when $[K/\sigma_K]_\oplus\gtrsim5$ (e.g., for our results using the Mat\'ern-5/2 activity kernel).
Future studies that account for realistic instrumental noise contributions in simulations such as those discussed in Section \ref{sec:fisher_simulations} are strongly encouraged.


\subsection{Stellar Mass Dependence}\label{sec:fisher_stellar_mass}

\begin{figure*} 
    \centering
    \includegraphics[width=1.0\linewidth]{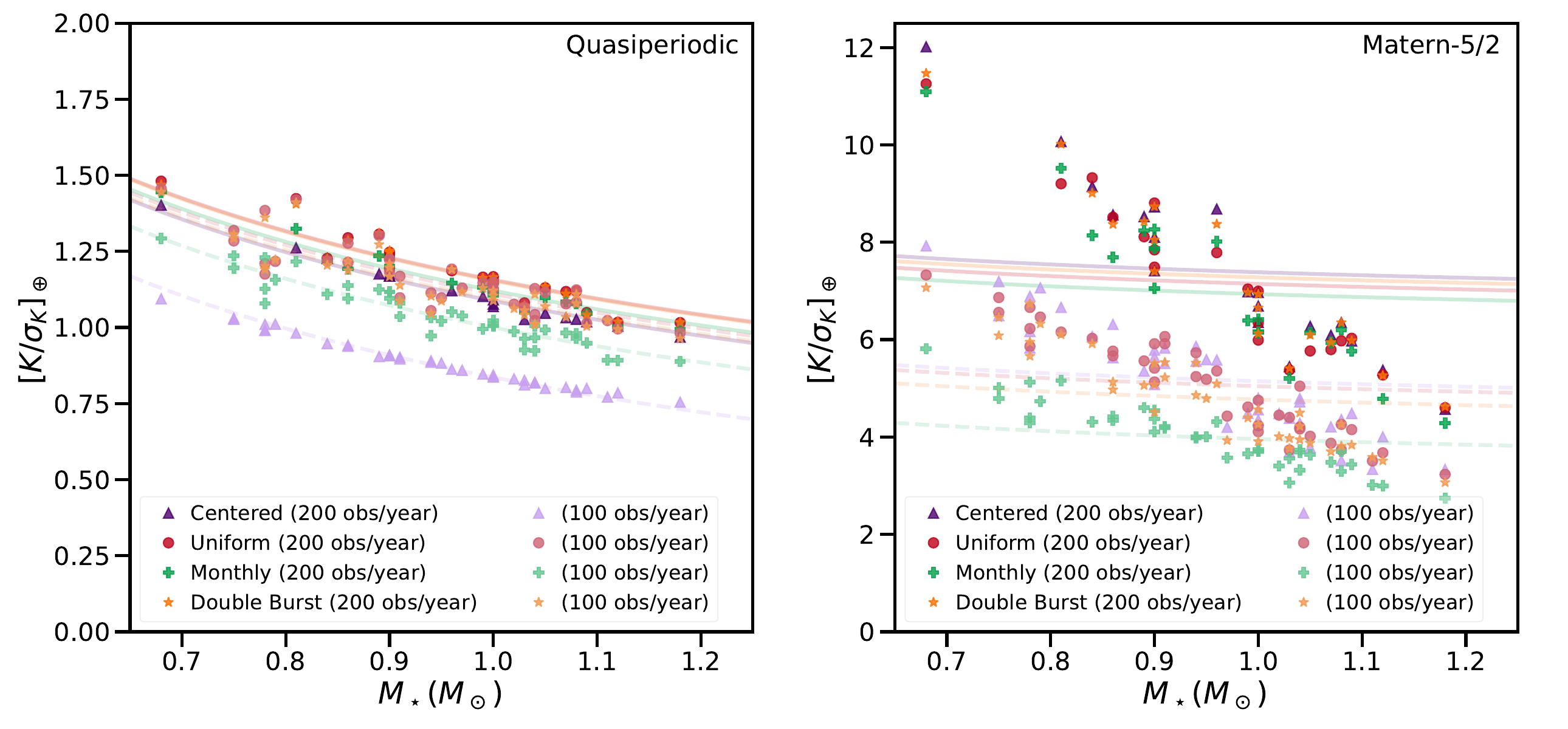}
    \caption{Dependence of expected Earth analog detection significance on stellar mass for each of our simulated surveys. On the left, we show that for the quasiperiodic kernel, $[K/\sigma_K]_\oplus$ follows a $K\propto M_\star^{-2/3}$ curve for every strategy. That is, $\sigma_K$ is independent of stellar mass. On the right, we show that the $[K/\sigma_K]_\oplus$ has a much steeper dependence on $M_\star$, suggesting that $\sigma_K$ depends on stellar mass as well.}
    \label{fig:fisher_mass_scaling}
\end{figure*}

\rrev{Survey performance will vary from star to star as a function of stellar mass, and we expect to achieve better sensitivity to exoplanets orbiting low mass stars given that $K\propto M_\star^{-2/3}$. As we show in \autoref{fig:fisher_mass_scaling}, this dependence is reflected in our results for the quasiperiodic activity kernel. The achieved detection significance, $[K/\sigma_K]_\oplus$, shows little scatter about a simple $M_\star^{-2/3}$ relation. However, this relation does not adequately explain the Mat\'ern-5/2 activity kernel detection significance results, which exhibit a much steeper mass dependence. This suggests that $\sigma_K$ depends on mass as well for this model.}
\citet{Luhn2023} see a similar trend in their survey analysis results and arrive at the conclusion that this is a consequence of the stellar parameter dependence of the oscillation and granulation correlated noise signals. Low-mass stars thus carry even more of an advantage for low-mass exoplanet detection than the RV semi-amplitude scaling would suggest. 
We note that this effect is only observed in the Mat\'ern-5/2 activity kernel results because the correlated noise contributions from the quasiperiodic activity kernel are so large that they mask any strong dependence on the asteroseismic correlated noise.

\subsection{Sensitivity to assumed definition of an Earth analog}\label{sec:fisher_analog_box}

\begin{figure*}
  \centering
  \includegraphics[width=\textwidth]{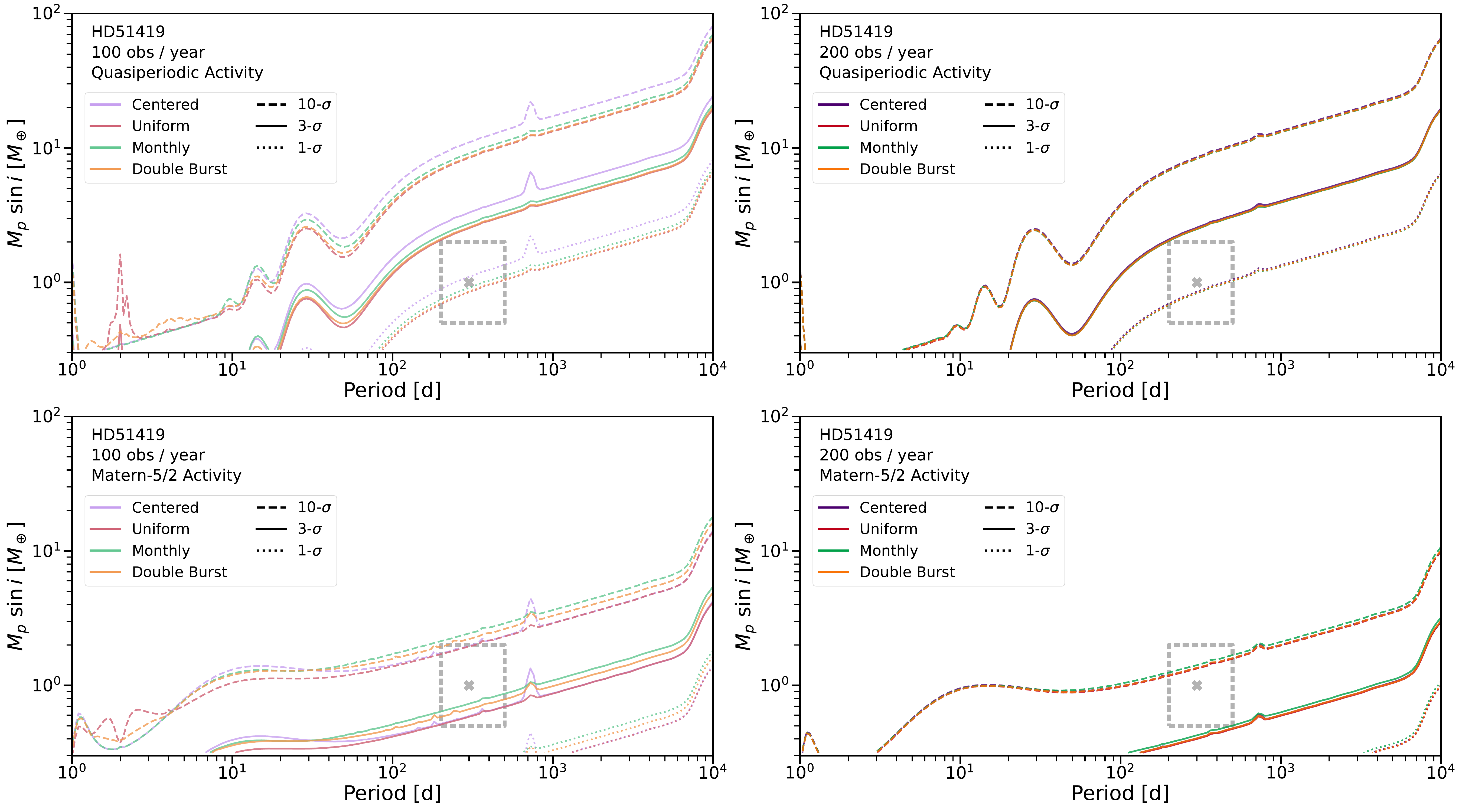}
\caption{Detection significance contours for a typical target star. For each simulated survey, we show the $K/\sigma_K=10$, 3, and 1 contours as solid, dotted, solid, and dashed lines, respectively. Certain features are more prominent for some survey strategies, such as the strong two-year alias peak for the Centered strategy and the lower sensitivity of the Uniform strategy to short-period planets. But at each significance level, it is clear that the local slope within the $D_\oplus$ Earth analog region (grey) does not change from strategy to strategy.}
\label{fig:fisher_contours}
\end{figure*}

Our definition of an Earth analog as applied to the $D_\oplus$ survey success metric is somewhat arbitrary and does not fully capture the commonly cited objective of ambitious EPRV surveys, which is to detect and measure the masses and orbits of Earth-mass planets in the Habitable Zones of their host stars \citep{Crass2021}. Because we fix the period range to 200 days $<P<$ 500 days for all stars, we do not account for the change in the location of the Habitable Zone as a function of stellar luminosity. Similar arguments can be made to change our mass limits of $0.5M_\oplus<M_p\sin i<2M_\oplus$.
However, we show in \autoref{fig:fisher_contours} that while certain features in the detection significance contours are more pronounced for some strategies than for others, the local slope of these contours over our specified period and mass ranges is quite insensitive to survey strategy. Instead, the differences between the success metrics calculated for each strategy result primarily from linear displacements of the detection significance levels. Changing the boundaries or even the shape of the region we use to calculate $D_\oplus$ will therefore have little to no effect on the relative performance of different survey strategies. 

\subsection{RV Model Complexity}\label{sec:fisher_model_complex}

In this work, we employ a \rrev{simple 3-parameter} model for the exoplanet-induced RV signal, assuming only one planet with a circular orbit. While our model allows us to focus on and compare specific survey strategy choices, future investigations in this vein could benefit from additional model complexity \rrev{(e.g., non-zero eccentricity or multiple planet signals)}.
Our approach also implicitly assumes perfect knowledge of $\gamma$, the absolute RV of the stellar rest frame, and of the RV signals imparted by any other exoplanets in the same system as the nominal Earth analog we are interested in detecting. In the case that $\gamma$ is the same for all observations, it can be accounted for by adding a constant offset term to \autoref{eq:rvmodel}, \rrev{which will have no effect on the resulting Fisher information}. But in practice, it is often necessary to combine data from multiple instruments to enable the detection of RV signals from low-mass, long-period exoplanets. The slight differences in the absolute RV scale for each instrument will necessitate the inclusion of instrument-specific $\gamma$ terms. A time-dependent $\gamma$ term can also be used to account for RV zero-point offsets in the data stream from a single instrument, such as those introduced by the SOPHIE octagonal fiber upgrade \citep{Bouchy2013} or the HIRES detector upgrade. Post-survey detection sensitivity analyses should take care to explicitly account for a varying $\gamma$ term.

The assumption that we will have detected and characterized all other significant exoplanet-induced signals in each system is also likely too optimistic. First, we note that several studies have found evidence for elevated co-occurrence rates for inner terrestrial planets and cool gas giants \citep{Bryan2019,Rosenthal2022} and multi-planet systems in general have been shown to be fairly common \citep{Fang2012,Zhu2022}. While EPRV surveys may easily detect and characterize a Jupiter analog with $K\sim 10$ m~s$^{-1}$, uncertainties on the mass and orbit will still complicate the detection of smaller RV signals. In addition, signals from multiple small planets with orbital periods close to 1 year, e.g., an Earth analog and a Venus analog, may be particularly difficult to disentangle. \rrev{Though it is beyond the scope of the present work,} expanding our RV model to account for signals from multiple planets within the Fisher information framework will enable us to determine the magnitude of this effect on Earth analog detection.

\section{Conclusions}\label{sec:fisher_conclusion}

We outline a detailed framework for analyzing the detection sensitivity of EPRV exoplanet surveys and we use this to assess the efficacy of various survey strategies in the correlated noise dominated regime. By using Fisher information analysis to quantify the impact of intrinsic stellar variability on exoplanet detection we can capture these correlated noise contributions with GPs and we \rrev{show that this method accurately reproduces the results of injection and recovery tests of synthesized} RV time series measurements.

We describe the design and execution of several tests with which we isolate and assess the impact of sources of noise on different timescales. From our simulations of survey schedules with different intra-night visit distributions, we show that:
\begin{itemize}
    \item distributed multi-visit strategies outperform single-visit strategies across all nightly time costs in the presence of granulation alone, but this advantage is reversed for oscillations
    \item in the oscillations-only case, one can achieve higher precision for visits shorter than $\lesssim20$ minutes by sampling the oscillation signal with a sequence of consecutive exposures instead of averaging over the signal with a single continuous exposure
    \item in the presence of correlated noise from oscillations and granulation together, single-visit strategies are generally preferred
    \item when we account for photon noise, multi-visit strategies are more heavily penalized by the precision loss incurred by overhead costs 
\end{itemize}
Analogous survey schedule simulations are used to test the performance of different distributions of observations across an observing season, and we find that:
\begin{itemize}
    \item dense, high-cadence observing strategies are preferred when rotationally modulated stellar activity signals are best represented by a Mat\'ern-5/2 kernel with a relatively short correlation length
    \item sparser observing strategies that maximize the seasonal observing baseline are preferred when these same signals are represented by a quasiperiodic kernel with a correlation length on par with the Solar rotation period
\end{itemize}


These findings are applied to simulations of 10-year EPRV exoplanet surveys with realistic target lists and time constraints as well as practical observing constraints such as weather losses, intra-survey competition, and seasonal observability. We calculate and compare the expected \rrev{parameter uncertainties} for each survey and show that the results of these more comprehensive simulations are largely consistent with those of the isolated tests. We also quantify the absolute detection sensitivity limits we will expect to achieve and discuss our findings. For the Mat\'ern-5/2 activity kernel, we expect to be sensitive Earth analog exoplanets around as many as $\sim 20$ quiet stars at the $[K/\sigma_K]_\oplus>5$ level for most survey strategies, whereas for the quasiperiodic kernel, we will achieve a typical detection significance of just $[K/\sigma_K]_\oplus>1$.

We also summarize several caveats to our results, the most important of which is that our analysis implicitly assumes perfect knowledge of the GP kernels with which we represent each source of stellar variability.
Our results show that the optimal survey strategy will depend strongly on the assumed hyperparameters and GP kernel form for certain sources of noise, suggesting that we should exercise caution in applying Fisher information analysis to specific stars for which these values are not well constrained. However if the kernels and hyperparameters are known to high accuracy and precision, the analysis framework we have outlined here can be a powerful tool with which we can identify more efficient and effective observing strategies both for individual stars as well as for an entire survey.

\section{Acknowledgements}

The authors would like to thank David W. Hogg for illuminating discussions on the application and interpretation of Fisher information analysis in this work, Jacob Luhn for his assistance with the oscillation and granulation GP kernel integrations, Christian Gilbertson for their guidance on the use of Cholesky factorization for matrix inversion and for discussions on the efficacy of different stellar activity kernels, and Sam Thompson and Annelies Mortier for feedback on historical weather patterns at La Palma. We appreciate the support and feedback from NEID science team members including Jason Wright, Suvrath Mahadevan, and Eric Ford as well as from \rrev{Christopher Lam and other} members of the Astronomical Data group at the Center for Computational Astrophysics as this project progressed. \rrev{We also thank the anonymous referee for constructive suggestions that greatly improved the final version of this manuscript.}

The Pennsylvania State University campuses are located on the original homelands of the Erie, Haudenosaunee (Seneca, Cayuga, Onondaga, Oneida, Mohawk, and Tuscarora), Lenape (Delaware Nation, Delaware Tribe, Stockbridge-Munsee), Shawnee (Absentee, Eastern, and Oklahoma), Susquehannock, and Wahzhazhe (Osage) Nations.  As a land grant institution, we acknowledge and honor the traditional caretakers of these lands and strive to understand and model their responsible stewardship. We also acknowledge the longer history of these lands and our place in that history.
The Center for Exoplanets and Habitable Worlds is supported by the Pennsylvania State University and the Eberly College of Science.
This research has made use of the SIMBAD database, operated at CDS, Strasbourg, France, and NASA's Astrophysics Data System Bibliographic Services.

\software{\textsf{astroplan} \citep{Morris2018},
\textsf{astropy} \citep{AstropyCollaboration2018}, \textsf{matplotlib} \citep{Hunter2007},
\textsf{numpy} \citep{Harris2020},
\rrev{\textsf{numpyro} \citep{Phan2019},}
\textsf{scipy} \citep{Oliphant2007}},
\rrev{\textsf{tinygp} \citep{Foreman-Mackey2024}}

\bibliography{references}{}
\bibliographystyle{aasjournal}

\appendix

\section{Survey Simulations}

\subsection{Target Selection}

The target list for our survey simulations is based on that of the NEID Earth Twin Survey (NETS), an ongoing EPRV exoplanet survey with the NEID spectrograph \citep{Schwab2016} on the WIYN-3.5m telescope at Kitt Peak National observatory. This list was assembled according to a set of quantitative prioritization metrics designed to identify stars conducive to EPRV Earth-analog searches \citep{Gupta2021} in conjunction with  observability considerations. The final NETS target list contains 41 stars, 40 of which are G- and K-dwarfs and one of which is an M-dwarf. We adopt the 40 Sun-like stars as our target sample for the survey simulations in this work; these stars are listed in \autoref{tab:fisher_targets}.
Although some of the NETS prioritization metrics are specific to the NEID spectrograph, we expect they are reasonably suitable for HARPS-3 given the similar spectral ranges over which the the instruments operate. In addition, these two spectrographs are at similar latitudes ($l_{\rm NEID} = 31^\circ57'$N, $l_{{\rm HARPS}-3} = 28^\circ46'$N), so constraints on target observability and intra-survey competition will not significantly differ.

To select targets for the 20-star subsample, we first sort the stars by the length of their observing season, which we calculate as the number of nights for which the star can be observed with HARPS-3 at an altitude of $>30^\circ$ (airmass $<$ 2) for at least 30 minutes between evening astronomical twilight and morning astronomical twilight. We select the top 20 stars, then substitute several stars with slightly shorter observing seasons to improve the on-sky distribution and reduce the potential for scheduling conflicts. Selecting stars with longer observing seasons ensures that we no not downplay differences between long-term survey strategies. In \autoref{tab:fisher_targets} we list the length of each star's observing season and we indicate which stars are included in the limited sample.

\subsection{Time Allocation}

To estimate the total time allocation for our simulated surveys, we first calculate the total annual observing time available to HARPS-3, which will be on the Isaac Newton Telescope on La Palma, Canary Islands. The mean duration of a night from this location, which we define as the time between evening astronomical twilight and morning astronomical twilight, is 9.1 hours. Assuming an annual weather loss of 16\% \citep[based on historical site statistics from La Palma;][]{Hall2018} and that an average of 2 nights per month will be reserved for instrument, telescope, and facility maintenance, the total observing time for HARPS-3 is then 2606 hours per year. Our assumed allocation will be half of this, or 1303 hours, which we round to 80,000 minutes. We assume a typical fixed time cost of 20 minutes per night each time a star is observed, which is sufficient for 100 nights per star for the 40-star sample and 200 nights per star for the 20-star sample.

In this regime the Fisher information criterion favors single-visit strategies, so for each night that a star is observed, we assign just one 20-minute visit. Each visit is scheduled as a sequence of three exposures with $t_{\rm exp} = 320$ seconds each, and we set $t_r = 30$ seconds and $t_{\rm acq} = 180$ seconds for a total cost of $t_{\rm night}=20$ minutes as calculated via \autoref{eq:tnight}. For the handful of stars in our sample brighter than $V<4.68$, we maintain the same total time cost but divide each visit into a sequence of six exposures with $t_{\rm exp} = 145$ seconds each. 
This is the magnitude threshold at which a 320 second exposure of a $T_{\rm eff} = 5500$ K star would saturate the detector assuming the same cross-dispersion spread and well depth as NEID (see Section \ref{sec:sim_covar_phot}).

\subsection{Scheduling Framework}

To construct our simulated surveys, we start by building a 10-year series of nightly observing schedules. We assume no prior knowledge of weather losses or other constraints on when observations can be scheduled, but binary masks are used to delineate allowed and disallowed nights for each star to approximate the cadences of each of the inter-night survey strategies we are testing. The number of allowed nights is set to be 20--30\% greater than the intended number of visits each year (e.g., $\sim125$ allowed nights per star for the 40-star sample and $\sim250$ allowed nights per star for the 20-star sample), such that each star is still observed close to the intended number of times after losses have been accounted for.

\subsubsection{Nightly Schedules}

For each night, we schedule observations for the set of allowed targets in order of increasing time of peak altitude. That is, the first star to be added to the schedule will be the star that reaches its highest altitude closest to evening astronomical twilight. If two stars reach their peak at the same time (e.g., stars that are already setting at twilight), priority is given to the star that is lower in the sky at this time. Observations are scheduled at the time of peak altitude. We record the start time and duration of each exposure, and a 180 second acquisition time is blocked off at the end of each sequence so that each observation consists of a contiguous 20 minute block. Any subsequent observation that would overlap with one that was previously scheduled is shifted in time until the start time of the observation coincides with the end of the previous block. 
We continue to add observations to the schedule in this manner until an observation crosses midnight, at which point we repeat the process in reverse, starting with morning astronomical twilight and working backwards until we again hit midnight.
If the start time of an observation would be shifted by more than 3 hours from the time of peak altitude or if the airmass at the time of observation would exceed 2, the observation will not be scheduled on that night.

\subsubsection{Simulating Losses}

After the schedule is generated, we mask a fraction of the nights to account for anticipated observing losses.

\begin{itemize}
    \item \textbf{Seeing: }In poor seeing conditions, guiding errors will introduce additional RV uncertainty and the flux captured by the fiber will decrease, in some cases making the data unsuitable for high precision RV measurements. To simulate seeing losses, we na\"ively assume seeing is constant throughout each night and we draw values from the following Rayleigh distribution 
\begin{equation}\label{eq:seeing}
    \begin{split}
    P({\rm FWHM};\sigma) &= \frac{{\rm FWHM}}{\sigma^2}e^{-{\rm FWHM}^2/(2\sigma^2)} +0.5\\
    \sigma &= 0.68
    \end{split}
\end{equation}
with a median full-width half-max (FWHM) of $1.3''$ to match empirical conditions at the Isaac Newton Telescope \citep{Thompson2016} and a minimum of $0.5''$. We set an upper tolerance on the seeing FWHM of $2.5''$ and mask all nights for which the seeing exceeds this limit.
\item \textbf{Weather losses: }Taking the binary mask constructed by \citet{Hall2018} based on historic weather losses at La Palma, we calculate the average fraction of nights lost to weather during each calendar month, $f_{m}$. For each scheduled night, we then compare the corresponding $f_{m}$ value to a randomly generated number $f_w$ on the interval $0\leq f_w < 1$. The following criteria are used to determine which parts of each night to mask, if any:
\begin{itemize}
    \item[] $0 \leq f_w < \frac{1}{2}f_{m} \Longrightarrow$ full night masked
    \item[] $\frac{1}{2}f_{m} \leq f_w < f_{m} \Longrightarrow$ first half masked
    \item[] $f_{m} < f_w \leq \frac{3}{2}f_{m} \Longrightarrow$  last half masked
    \item[] $\frac{3}{2}f_m \leq f_w < 1 \Longrightarrow$ full night open.
\end{itemize}
In practice, this adds some granularity to the weather losses without changing the average time we would lose per month if we simply masked all nights for which $0\leq f_w < f_m$.
\item \textbf{Maintenance: }We mask two randomly selected nights out of every month to simulate a telescope and facility maintenance schedule.
\end{itemize}

\begin{figure*} 
    \centering
    \includegraphics[width=1.0\linewidth]{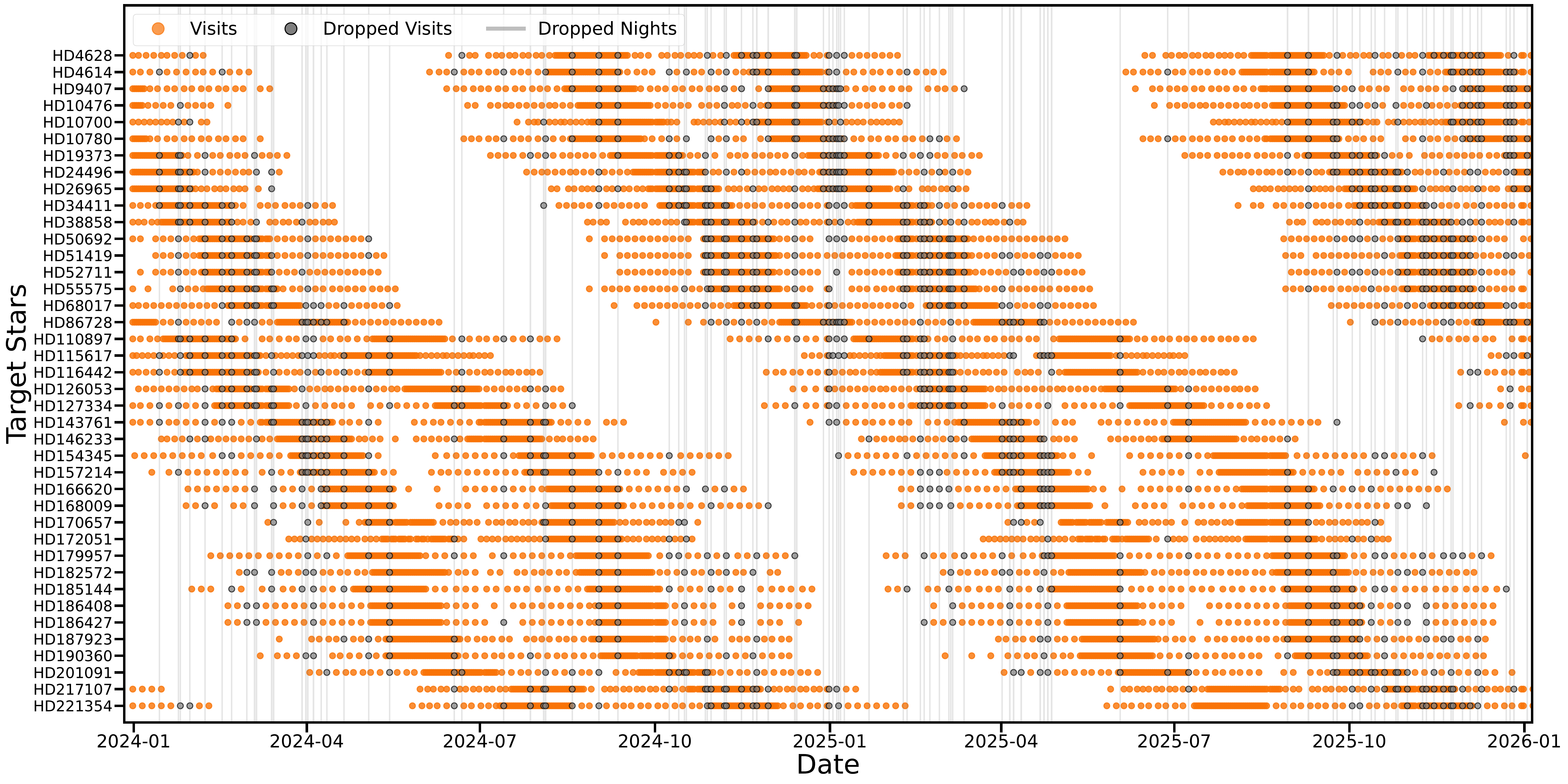}
    \caption{First two years of simulated observations for the Double Burst inter-night observing strategy as applied to the 40-star target sample with practical constraints on target observability. Stars are shown in order of increasing RA from top to bottom. Orange markers represent nights during which each target was observed, grey markers show observations that were scheduled and then dropped due to simulated weather losses and telescope closures, and grey lines indicate the nights on which these losses occurred. For most stars, we are able to achieve a reasonable approximation of the idealized version of this strategy shown in \autoref{fig:fisher_spotstrat} in spite of these practical constraints.}
    \label{fig:fisher_schedule}
\end{figure*}

All observations in any of these masked windows are dropped from the schedule; we note that these masks are not expected to be mutually exclusive, so they are drawn independently and will overlap in some cases. In \autoref{fig:fisher_schedule}, we show the first two years of observations and anticipated losses for one of our simulated survey schedules.

\end{document}